\documentclass[12pt]{iopart}
\usepackage{graphicx}
\usepackage{wrapfig, framed, caption}
\usepackage{braket}
\usepackage{dsfont}
\usepackage{float}
\usepackage{cite}
\usepackage{bm}
\usepackage{xcolor}

\pdfoutput=1

\usepackage{iopams}  
\begin{document}

\title{Growing macroscopic superposition states via cavity quantum optomechanics}

\author{Jack Clarke$^{1,2}$ and Michael R. Vanner$^{1,2}$}

\address{$^1$QOLS, Blackett Laboratory, Imperial College London, London SW7~2AZ, UK}
\address{$^2$Clarendon Laboratory, Department of Physics, University of Oxford OX1~3PU, UK}
\ead{jack.clarke@imperial.ac.uk, m.vanner@imperial.ac.uk}

\begin{abstract}
The investigation of macroscopic quantum phenomena is a current active area of research that offers significant promise to advance the forefronts of both fundamental and applied quantum science. Utilizing the exquisite precision and control of quantum optics provides a powerful toolset for generating such quantum states where the types and `size' of the states that can be generated are set by the experimental parameter regime available and the resourcefulness of the protocol applied. In this work we present a new multistep scheme to `grow' macroscopic superposition states of motion of a mechanical oscillator via cavity quantum optomechanics. The scheme consists of a series of optical pulses interacting with a mechanical mode via radiation-pressure followed by photon-counting measurements. The multistep nature of our protocol allows macroscopic superposition states to be prepared with a relaxed requirement for the single-photon optomechanical coupling strength. To illustrate the experimental feasibility of our proposal, we quantify how initial mechanical thermal occupation and mechanical decoherence affects the non-classicality and macroscopicity of the states generated and show that our scheme is resilient to optical loss. The advantages of this protocol provide a promising path to grow non-classical mechanical quantum states to a macroscopic scale under realistic experimental conditions.
\end{abstract}

\vfill
\section{Introduction}

Studying macroscopic quantum states has a myriad of motivations that range from quantum technology development to deepening our understanding of the foundations of physics. Notably, observing the dynamics of such states can put tighter bounds on potential models for wavefunction collapse~\cite{GRW1986,Penrose1996,Diosi1989}, and provides a path to test macroscopic quantum phenomena and quantum gravity on a table top~\cite{BJK1999,Marshall2003,Pikovski2012,Bosso2017,Bose2017,Marletto2017,Bekenstein2012}. Current experimental platforms pursuing the preparation of macroscopic quantum states include molecule interferometry~\cite{Haslinger2013}, superconducting circuits~\cite{Clarke2008,Devoret2013}, ultracold atoms~\cite{Berrada2013}, and cavity quantum optomechanical systems~\cite{RMP2014}.

The present work contributes to cavity quantum optomechanics, which utilizes optical forces and the quantum control of light to generate and study non-classical states of mechanical motion. The field has diversified significantly over the last two decades and a number of quite different experimental platforms and protocols are now being explored. Prominent deterministic protocols to generate non-classical mechanical states include the generation of mechanics-field entanglement~\cite{BJK1999,Marshall2003,Armour2002}, and non-Gaussian state preparation via optomechanical state-swap~\cite{Akram2010,Khalili2010,Marek2010,Bennett2016}. Utilizing measurement provides a powerful non-deterministic approach for state preparation with prominent examples including mechanical squeezing via measurement~\cite{Vanner2011,Lei2016}, superposition state preparation via position-squared measurements~\cite{Romero2011,VannerPRX,Brawley2016}, phonon addition/subtraction~\cite{Lee2012,Vanner2013,Riedinger2016}, and sequential measurement schemes~\cite{Milburn2016,Hoff2016}. This line of research lays the foundation for the development of optomechanical quantum technologies such as microwave-to-optical conversion~\cite{Andrews2014,Tian2015,Xia2014}, weak force sensing~\cite{Rugar2004,Hosseini2014}, and quantum information applications~\cite{Stannigel2012}, by establishing and improving the quantum coherence of mechanical motion.

Here we introduce and theoretically develop a versatile new scheme for non-classical mechanical state preparation via measurement that can `grow' a mechanical superposition state with a sequence of optical pulsed interactions and photon-counting measurements. Our scheme operates outside the resolved-sideband regime, and builds upon the operation introduced in Ref.~\cite{Ringbauer2016}, where a single pulsed interaction and photon-counting measurement causes the mechanical oscillator to undergo a superposition of the identity operation and a momentum kick. For this operation, the momentum transfer is determined by the single-photon optomechanical coupling strength. At present, this single-photon coupling strength is small for experimental solid-state optomechanical systems, and hence our scheme provides a path to increase the momentum transfer by utilizing a sequence of many operations. As will be detailed below, this is achieved by appropriately choosing the phases in a sequence to cancel all the possible momentum components apart from the two extrema. In this way, we generate a quantum superposition of two well-separated mechanical momentum components, which comprise two macroscopically distinguishable states often referred to as a Schr\"{o}dinger cat state. In contrast to existing schemes in the literature for superposition state generation, our approach has several advantages: (i) the requirement for strong single-photon optomechanical coupling is relaxed and larger superposition states can be generated by making more steps, (ii) only easily prepared optical inputs states, such as coherent states and single photons, are required, (iii) the scheme is resilient to optical loss, and (iv) an avenue is opened for further studies with multiple operations to generate a wide variety of superpositions in momenta as well as throughout mechanical phase-space. In this work we focus on the preparation of mechanical Schr\"{o}dinger cat states, due to their importance in many proposals for exploring the limits of standard quantum theory and for their utility in sensing and quantum information applications. Using a range of parameter sets from recent experiments, we quantify and illustrate the performance of our scheme by characterizing the non-classicality and macroscopicity of the mechanical states that can be generated. Even in the presence of mechanical thermal occupation and decoherence, we find that strong non-classicality, as indicated by the minimum of the Wigner function and its total negative volume~\cite{Kenfack2004}, and large macroscopicity, as defined via Ref.~\cite{Lee2011} and the quantum Fisher information~\cite{Frowis2018}, can be readily generated. Furthermore, successful state preparation can be performed with an experimentally reasonable heralding probability. Alongside this we also propose parameter sets for improved performance and explore the behaviour of the states produced in near-future realizations of our multistep scheme.

\section{Multistep protocol}
\subsection{Growing mechanical superposition states with a sequence of operations}\label{basicprotocol}
\begin{figure}
\begin{framed}\centering
\includegraphics[scale=0.5, angle=270]{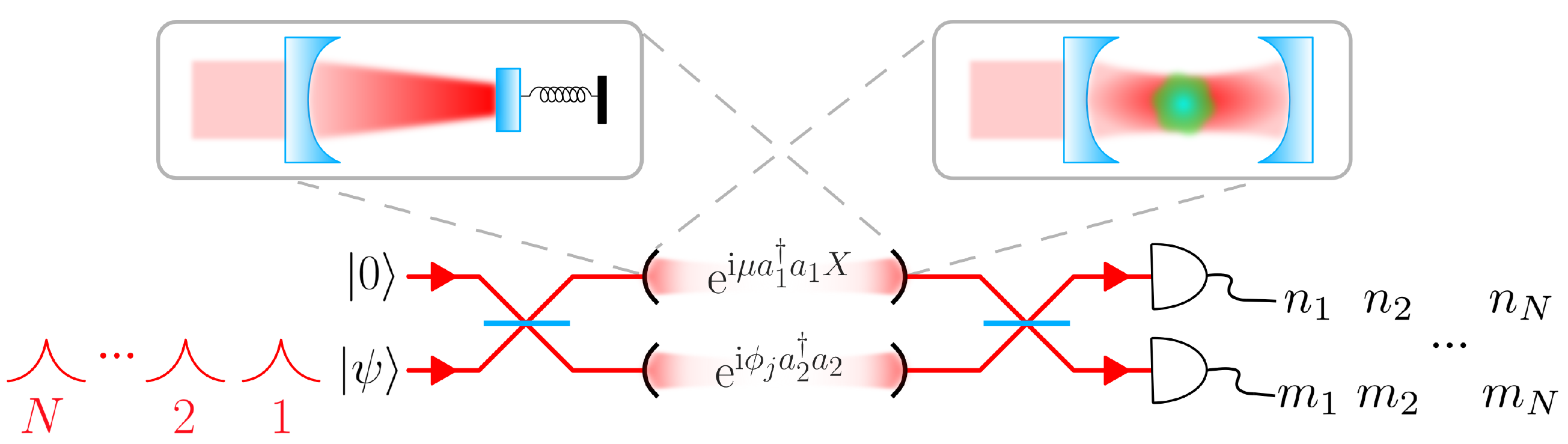} 
\caption{\small{{\textbf{Multistep optomechanical scheme for macroscopic superposition state preparation}}. $N$ optical pulses are sent through the interferometer that heralds the generation of a mechanical Schr\"{o}dinger cat state when a sequence of $\{0,1\}$, or $\{1,0\}$, click events are registered at the output. Upper left: cartoon of a solid-state cavity quantum optomechanical system. Here, one of the end mirrors of a Fabry-P\'{e}rot cavity is suspended  and couples via radiation pressure to the intracavity optical field. Upper right: cartoon of a cavity quantum optomechanical device using ultracold atoms. Here, the density excitations of a Bose-Einstein condensate act as the mechanical oscillations and couple to the cavity field through the optomechanical interaction, equivalent to the solid-state system shown on the left.}}
\label{setuppdf}
\end{framed}
\end{figure}

Our optomechanical protocol for non-classical mechanical state preparation can be applied to a wide range of systems. In particular, the two platforms we consider in detail for the present work are engineered solid-state mechanical systems~\cite{RMP2014}, and ultracold atom implementations~\cite{Brennecke2008, Purdy2010}, where a cloud of atoms `sloshes' within a trapping potential. Our setup, see Fig.~\ref{setuppdf}, allows for the implementation of a multistep protocol, in which each step involves an optical pulse interacting with the mechanical mode of interest via radiation-pressure. The end of each step is affirmed by a photon-number measurement on the optical field, which heralds a nonlinear operation applied to the mechanical oscillator. In this section we will first describe how a single step of this protocol creates a superposition of mechanical momentum states. After this we will then describe how multiple steps may be used to enhance the separation in momentum and therefore `grow' mechanical superposition states. Before the protocol is outlined, it is convenient to introduce some of our notation. Firstly, $X=(b+b^{\dagger})/\sqrt{2}$ is the mechanical position operator in units of the zero-point motion of the mechanics $x_{0}=\sqrt{\hbar/(m\omega)}$, where $\omega$ is the mechanical frequency and $m$ is the effective mechanical mass. Secondly, $P=-\rmi(b-b^{\dagger})/\sqrt{2}$ is the mechanical momentum operator in units of the zero-point momentum, $p_{0}=\sqrt{\hbar{m}\omega}$. When $X$ and $P$ are defined in this way we have $[X,P]=\rmi$.\par

Each step of our protocol is initiated by the injection of a pulse of light in the quantum state $\ket{\psi}$ into a Mach-Zehnder interferometer via a 50:50 beam splitter. For this scheme, this optical pulse is taken to be either a weak coherent state ${\ket{\sqrt{2}\alpha}}$ or a single photon $\ket{1}$. We will describe both cases here and show that, apart from the heralding probability, both of these optical input states result in the same non-unitary operation, which generates the mechanical superposition of momenta. At the first beam splitter, the optical pulse is mixed with vacuum on the unused port through $U^{\dagger}a_{1}U=(a_{1}+a_{2})/\sqrt{2}$ and $U^{\dagger}a_{2}U=(a_{1}-a_{2})/\sqrt{2}$, where $a_{1,2}$ are the annihilation operators for the two optical modes of the interferometer. In the lower arm, a phase $\phi_{j}$ is imprinted onto the optical field, where $j$ indexes the step number. While in the upper arm the optical pulse and mechanical mode couple via the interaction Hamiltonian
 \begin{equation}
H_{\mathrm{int}}=-\hbar{g}_{0}a_{1}^{\dagger}a_{1}(b+b^{\dagger}),
\end{equation}
where $g_{0}$ is the optomechanical coupling rate. Provided that the interaction time is much shorter than the mechanical period, we model the optomechanical interaction with the unitary operation $\mathrm{e}^{\rmi{\mu}a_{1}^{\dagger}a_{1}X}$, as has been employed in Ref.~\cite{Vanner2011}. Here, $\mu$ is the dimensionless optomechanical coupling strength, which quantifies the momentum transfer during the interaction in units of zero-point momentum. For a single-sided optomechanical cavity $\mu=3g_{0}/\sqrt{2}\kappa$, where $\kappa$ is the cavity amplitude decay rate. The numerical prefactor in $\mu$ originates from the assumed temporal shape of the optical pulse, $\sqrt{\kappa}\exp(-\kappa|t|)$, which matches the spectrum of the cavity~\cite{Vanner2011}, and a derivation of $\mu$ in this case is given in \ref{pulseshapeappendix}. For the case of the coherent input, this temporal envelope may be engineered from a continuous laser by amplitude modulation. Whereas for single photons, this pulse shape is often created naturally in the process of cavity-enhanced non-degenerate parametric downconversion~\cite{Neergaard-Nielsen2007}. After this optomechanical interaction, the two optical modes then interfere on a second 50:50 beam splitter and photon-number measurements are made at the beam-splitter outputs, which provides the event-ready signal for the end of the step. Temporal mode matching of these two optical pulses may be ensured by placing a cavity, with an identical response function to that of the optomechanical cavity, in the bottom arm of the interferometer. An $\{m_{j},n_{j}\}$ click event corresponds to detecting $m$ photons in the detector at the output of mode $1$, and $n$ photons at the output of mode $2$ at the end of the $j^{\mathrm{th}}$ step.\par

 \par

Registering an $\{m_{j},n_{j}\}$ click event for a single step of the protocol corresponds to a measurement operator that maps the initial state of the mechanical mode $\rho_{\mathrm{in}}$ to the output state $\rho_{\mathrm{out}}\propto\Upsilon^{(j)}_{m_{j},n_{j}}\circ\rho_{\mathrm{in}}$. For brevity we use the circle notation for quantum operations, i.e. $\Upsilon\circ\rho=\Upsilon\rho{\Upsilon}^{\dagger}$. Here, the superscript $j$ on the measurement operator highlights the dependence on the choice of phase at the $j^{\mathrm{th}}$ step, while the subscripts run over the complete set to which the measurement operator $\Upsilon^{(j)}_{m_{j},n_{j}}$ belongs. This measurement operator is given by $\Upsilon_{m_{j},n_{j}}^{(j)}=\bra{m_{j}}\bra{n_{j}}U\mathrm{e}^{\rmi{\mu}a_{1}^{\dagger}a_{1}X}\mathrm{e}^{\rmi{\phi_{j}}a_{2}^{\dagger}a_{2}}U\ket{\psi}\ket{0}$, where $U$ is the beam splitter operation defined above. More explicitly, we have 
\begin{equation}\label{kraussoperator}
\fl \begin{array}{l} 
 \Upsilon_{m_{j},n_{j}}^{(j)}=\left\{
    \begin{array}{l}
      \frac{{\rm{e}}^{-|\alpha|^2}}{\sqrt{m_{j}!n_{j}!}}\frac{\alpha^{m_{j}+n_{j}}}{(\sqrt{2})^{m_{j}+n_{j}}}({\rm{e}}^{\rmi\mu X}+{\rm{e}}^{\rmi\phi_{j}}\mathds{1})^{m_{j}}({\rm{e}}^{\rmi\mu X}-{\rm{e}}^{\rmi\phi_{j}}\mathds{1})^{n_{j}}\quad\quad\hspace{0.05cm} \mathrm{for}~{\ket{\psi}}={\ket{\sqrt{2}\alpha}}\smallskip \\
      \frac{1}{2}[({\rm{e}}^{\rmi\mu{X}}+{\rm{e}}^{\rmi\phi_{j}}\mathds{1})\delta_{m_{j},1}\delta_{n_{j},0}+({\rm{e}}^{\rmi\mu{X}}-{\rm{e}}^{\rmi\phi_{j}}\mathds{1})\delta_{m_{j},0}\delta_{n_{j},1}]\quad \mathrm{for}~{\ket{\psi}}={\ket{1}}.
    \end{array}
  \right.
  \end{array}
\end{equation}
It is thus seen that this operation applies a combination of displacement operators $\mathrm{e}^{\rmi\mu{X}}$ and phase shifts ${\rm{e}}^{\rmi\phi_{j}}\mathds{1}$ to the mechanical state dependent on the choice of phase and the measurement outcome.

Multiple steps can be used to enhance the separation size of these superposition states and, in this way, grow the quantum state to a macroscopic scale. To ensure that the protocol grows the superposition state along the momentum axis of phase space, the pulses are applied either one after another rapidly, in a time much less than the mechanical period, or once every mechanical period. In the following discussions, we will assume that the latter approach is adopted and in later sections compute the decoherence between steps. After $N$ steps of this state preparation protocol the mechanical density operator is described by
\begin{equation} \label{multimeas}
\rho_{N}=\frac{1}{P_{N}}(\Upsilon^{(N)}_{m_{N},n_{N}}...\Upsilon^{(2)}_{m_{2},n_{2}}\Upsilon^{(1)}_{m_{1},n_{1}}\circ\rho_{\mathrm{in}}).
\end{equation}
Here, the probability to obtain a particular series of $N$ output click events is given by $P_{N}=\mathrm{tr}(\Upsilon^{(N)}_{m_{N},n_{N}}...\Upsilon^{(2)}_{m_{2},n_{2}}\Upsilon^{(1)}_{m_{1},n_{1}}\circ\rho_{\mathrm{in}})$.  \par
For the particular case of a series of $\{0,1\}$ click events, every measurement operator in Eq.~(\ref{multimeas}) takes the form $\Upsilon_{0,1}^{(j)}
\propto({\rm{e}}^{\rmi\mu{X}}-{\rm{e}}^{\rmi\phi_{j}}\mathds{1})$, which is a superposition of a momentum displacement operator and a phase shift. In this case, the multistep protocol then maps the initial state $\rho_{\mathrm{in}}$ of the mechanics to the final state 
\begin{equation} \label{rhoout1}
\rho_{N}\propto\prod_{j=1}^{N}\Big[D(\rmi\mu/\sqrt{2})-{\rm{e}}^{\rmi\phi_{j}}\mathds{1}\Big]\circ\rho_{\mathrm{in}}.
\end{equation} 
The multistep protocol therefore produces a final mechanical state consisting of $N+1$ copies of the initial state along the momentum axis of phase space, each separated by $\mu$. Controlling the value of $\phi_{j}$ at each step of the protocol allows for the preparation of a variety of mechanical momentum superposition states. An especially noteworthy choice of phase is $\phi_{j}=2\pi{j}/N$, which leads to cancellation of all the cross terms in Eq.~(\ref{rhoout1}), leaving
\begin{equation} \label{rhoout2}
\rho_{N}\propto[D(\rmi{N}\mu/\sqrt{2})-\mathds{1}]\circ\rho_{\mathrm{in}}.
\end{equation}
Eq.~(\ref{rhoout2}) shows that the final state of the mechanics comprises a Schr\"{o}dinger cat state (SCS) along the momentum axis of phase space consisting of a superposition of the initial state at $P=0$ and $P=N\mu$. When this particular choice of phase is utilized the multistep protocol therefore leads to an enhancement in the phase-space separation of momentum components by a factor $N$. Growing the superposition state in this way will lead to more prominent non-classicality and macroscopicity features. As stated earlier, the SCS is frequently discussed in theoretical proposals that study quantum mechanical phenomena such as wavefunction collapse and the interface between quantum mechanics and gravity, and hence we focus on this class of state for the rest of this work. \par 
For further calculations, it is convenient to express and visualize these states in a phase space formed by position and momentum operators, $X$ and $P$, using the Wigner quasiprobability distribution. Assuming that the mechanical mode starts off in an initial thermal state with a mean thermal occupation $\bar{n}$, i.e.  $\rho_{\bar{n}}=(\pi\bar{n})^{-1}\int\mathrm{d}^{2}\beta\mathrm{e}^{-|\beta|^{2}/\bar{n}}\ket{\beta}\bra{\beta}$, where the dummy variable $\beta$ is a coherent amplitude, the final state of the mechanical mode after the multistep protocol given by Eq.~(\ref{rhoout2}) is described by the Wigner distribution  
\begin{eqnarray}
\fl \quad\quad W_{\mathrm{SCS}}(X,P)=\frac{{\mathcal{N}}_{\mathrm{SCS}}}{\pi(1+2\bar{n})}\Bigg\{&{\exp}\Big[\frac{-X^2-P^2}{1+2\bar{n}}\Big]+{\exp}\Big[\frac{-X^{2}-(P-N{\mu})^{2}}{1+2\bar{n}}\Big]\nonumber\\
&-2{\rm{cos}}(N{\mu}X){\exp}\Big[\frac{-X^2-(P-N{\mu}/2)^2}{1+2\bar{n}}\Big]\Bigg\} . \label{Wigscseq}
\end{eqnarray} 
Here, $1/2{\mathcal{N}}_{\mathrm{SCS}}= 1-\exp[-N^2\mu^{2}(1+2\bar{n})/4]$ ensures normalization. The first and second terms inside the curly brackets of Eq.~(\ref{Wigscseq}) are the population components of the SCS, while the last term corresponds to the quantum interference between them.
\begin{figure} 
\begin{framed}\centering
\includegraphics[scale=0.255,angle=270,trim=0 1.25cm 0 0]{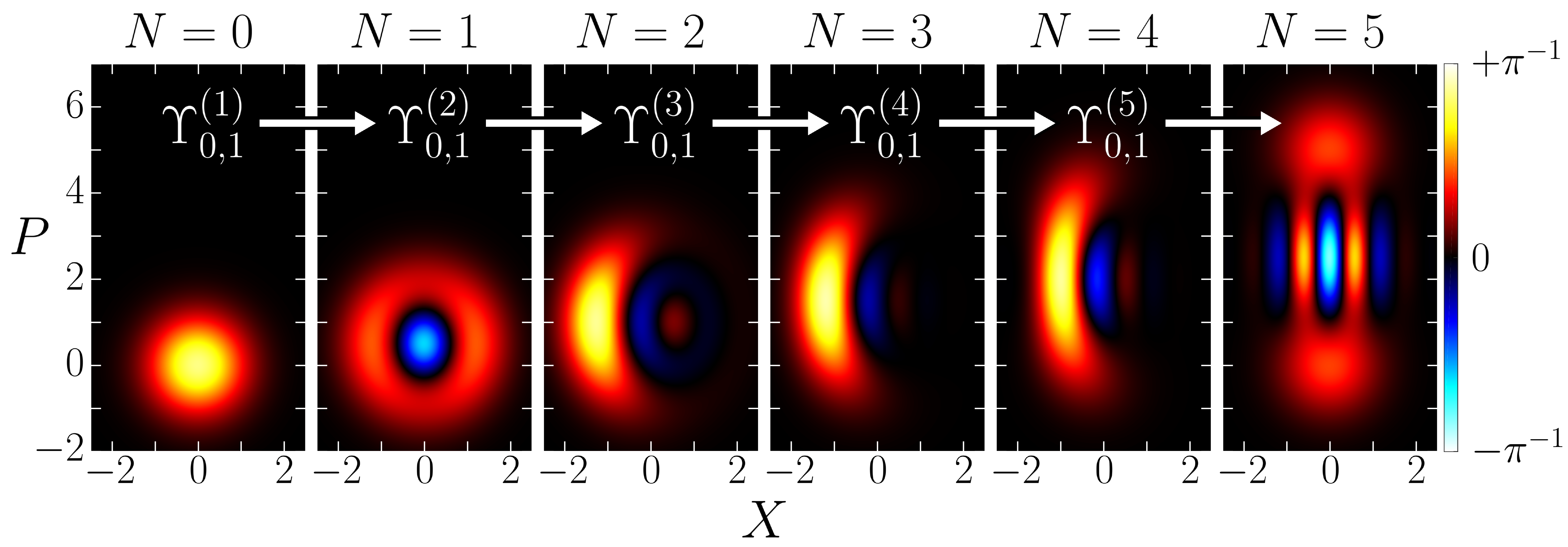} 
{\color{black}\caption{\small{{\textbf{Growing mechanical Schr\"{o}dinger cat states in a multistep process}}. Here, the Wigner distribution is plotted at each step of a five-step protocol for $\mu=1$ and $\bar{n}=0.1$. Between each plot the measurement operator $\Upsilon_{0,1}^{(j)}$ is applied to the previous mechanical state. The phases applied on the first, second, third, fourth and fifth steps are $\phi_{1}=0$, $\phi_{2}=2\pi/5$, $\phi_{3}=4\pi/5$, $\phi_{4}=6\pi/5$ and $\phi_{5}=8\pi/5$, respectively. The final plot of this figure is the SCS, consisting of two population terms centred at $P=0$ and $P=5$, with interference between these two terms centred at $P=2.5$.}}
\label{growingfig}}
\end{framed}
\end{figure} \par

This type of mechanical superposition state may also be obtained by observing a series of $\{1,0\}$ click events instead of a series of $\{0,1\}$. With $\phi_{j}=2\pi{j}/N$, registering a series of $\{1,0\}$ clicks leads to the generation of the final state $\rho'_{N}=\mathcal{N'}[D(\rmi\mu/\sqrt{2})^{N}-(-1)^N\mathds{1}]\circ\rho_{\mathrm{in}}$. The mechanical mode is therefore in an even cat state or an odd cat state if $N$ is odd or even, respectively. At small values of $N\mu$ the total measurement operator for the protocol is proportional to the identity operator when $N$ is odd, i.e. $D(\rmi\mu/\sqrt{2})^{N}+\mathds{1}\approx2\mathds{1}$. While for even $N$, at these small values of $N\mu$ the measurement operator is proportional to the position operator, i.e. $D(\rmi\mu/\sqrt{2})^{N}-\mathds{1}\approx{\rmi}N\mu{X}$. It is therefore apparent that a protocol which produces odd cat states for all $N$ is preferred, as it leads to the production of mechanical states which differ significantly from the initial mechanical state, even for small $\mu$. Crucially, these odd cat states possess more prominent non-classicality and macroscopicity features as will be discussed in Section \ref{non-classicality} and \ref{macroscopicity}. Therefore, to obtain these more interesting states for a series of $\{1,0\}$ click events at any $N$, and recover Eq.~(\ref{rhoout2}), the phase is chosen as $\phi_{j}=2\pi{j}/N+\pi$. In summary, the multistep protocol for growing a mechanical SCS consists of recording a series of $\{0,1\}$ or $\{1,0\}$ click events and choosing the phases to be $\phi_{j}=2\pi{j}/N$ or $\phi_{j}=2\pi{j}/N+\pi$, respectively. Fig.~\ref{growingfig} shows how the mechanical SCS grows in phase-space during a five step process.

\subsection{Heralding probability and optical loss}\label{Heraldingsection}
For no optical loss, the heralding probability $P_{N}$ for the multistep state-preparation protocol may be calculated by setting the trace of Eq.~(\ref{multimeas}) equal to one. To model detector inefficiency and optical loss, we insert fictitious beam splitters of intensity transmission $\eta$ after the cavities in the upper and lower arms of the interferometer in Fig.~\ref{setuppdf}. If there is an asymmetry between the optical loss in the upper and lower arms of the interferometer this may be compensated for by changing the transmission coefficient of the first beam splitter to balance the amplitudes at the final beam splitter. On these fictitious beam splitters, the optical modes of the interferometer then interact with the environment, which at optical frequencies is well described by the vacuum state, and a trace over the output environmental states is performed to account for the loss of optical information. For a single photon input state, optical losses only act to reduce the heralding probability for the multistep state-preparation protocol and do not affect the final mechanical state. This is because, when single photons are chosen as the input states, optical loss always prevents the click event which heralds the next step of the protocol and so that experimental run is discarded as explained below. Optical loss also has negligible effect on the final mechanical state when the input state is a weak coherent state, provided that $(1-\eta)|\alpha|^2\ll1$. When optical loss is significant, such that $(1-\eta)|\alpha|^2\ll1$ cannot be satisfied, the loss of photons to the environment leads to mechanical decoherence, which reduces the mechanical non-classicality. A more detailed treatment of optical loss is given in \ref{opticalloss}. Henceforth, in the main text we will assume that single photons or weak coherent states, satisfying the condition $(1-\eta)|\alpha|^2\ll1$, are used as the optical input states. These conditions ensure that optical-loss-induced mechanical decoherence is avoided and loss only acts to decrease the heralding probability for state preparation.

For an initial thermal state $\rho_{\mathrm{in}}=\rho_{\bar{n}}$, and using $\phi_{j}=2\pi{j}/N$, $m_{j}=0$, and $n_{j}=1$ for all $j$, we then have
\begin{equation} \label{heralding}
\fl P_{N}=\left\{
    \begin{array}{l}
      2^{1-N}\mathrm{e}^{-2N\eta|\alpha|^2}\eta^{N}|\alpha|^{2N}\Big\{1-\exp[-N^2\mu^2(1+2\bar{n})/4]\Big\}\quad  \mathrm{for}~{\ket{\psi}}={\ket{\sqrt{2}\alpha}} \\
      2^{1-N}\eta^{N}\Big\{1-\exp[-N^2\mu^2(1+2\bar{n})/4]\Big\}\quad \quad \quad \quad\quad\quad\,\,\, \mathrm{for}~{\ket{\psi}}={\ket{1}}.
    \end{array}
  \right.
\end{equation}
The case where ${\ket{\psi}}={\ket{1}}$ shows more favourable scaling in the heralding probability. This improved performance occurs for all coherent amplitudes, as can been seen explicitly if the first heralding probability in (\ref{heralding}) is maximized with respect to $\sqrt{\eta}|\alpha|$ to obtain $\sqrt{\eta}|\alpha|_{\mathrm{max}}=1/\sqrt{2}$, which maximizes the probability for $\{0,1\}$ and $\{1,0\}$ measurements, although smaller values of $|\alpha|$ may be needed to satisfy $(1-\eta)|\alpha|^2\ll1$ at realistic detector efficiencies, as discussed in \ref{opticalloss}. This difference in heralding probability is easily explained by photon number conservation, which demands that for a single photon input, and with no optical loss, the only possible output clicks are $\{0,1\}$ and $\{1,0\}$, which are optimal for the protocol described above. On the other hand, whereas the coherent input pulse permits $m_{j}$ and $n_{j}$ to take on any integer value greater than or equal to zero. We will now discuss realistic photon counting schemes for our two optical inputs. \par
For $ {\ket{\psi}}={\ket{1}}$, low-dark-count avalanche-photodiodes (APDs) may be employed. This is because, in the absence of loss, only $\{0,1\}$ or $\{1,0\}$ click events are possible, which are readily detected using APDs, and photon-number resolution is not required. In the presence of loss $\{0,0\}$ click events are possible, which leads to a loss of mechanical non-classicality and therefore runs of the experiment where these events occur are discarded. In this way, the protocol is made resilient to detector inefficiency and optical loss by selecting a successful run of the experiment at the cost of a reduced success probability. Dark counts present a second unavoidable deleterious effect as they introduce false positive measurement outcomes. To minimize the frequency of these events, the detection time window may be gated to the pulse arrival time. \par
When ${\ket{\psi}}={\ket{\sqrt{2}\alpha}}$, the protocol calls for high efficiency, low-noise, photon-number resolving detectors. As the number of steps in the protocol increases past $N=2$ then, depending on the experimental parameter regime, the total probability of obtaining the SCS may become smaller than the probability of observing a $\{0,2\}$, $\{2,0\}$ or $\{1,1\}$ at any particular step throughout the state preparation protocol. Therefore, in certain parameter regimes---such as $\mu<1$ and coherent amplitudes $|\alpha|\sim1$---if the detectors are not capable of photon-number resolution we cannot be confident that a true sequence of $\{0,1\}$ or $\{1,0\}$ events has been recorded and that optical loss has not led to a misidentification in the string of events, which would reduce the mechanical non-classicality. However, if we satisfy the above demands on detector performance---in particular that optical loss is low---then we may include $\{0,0\}$ click events within the string of click events and therefore boost the heralding probability for state generation. This is because the measurement operator for a $\{0,0\}$ event is proportional to identity and so leaves the mechanical mode unchanged when mechanical decoherence can be neglected. \par
Following our discussions of the heralding probability for state preparation, we may now offer an approximate timescale over which the experiment may be carried out. Suppose that input pulses are applied every period and the mechanics is allowed to relax during a time $T_{\mathrm{r}}$ between each run of the experiment, where a run involves $N$ steps of the protocol. Here, $T_{\mathrm{r}}=\min\{1/\gamma,2\times10^3\pi/\omega\}$, meaning that we let the mechanical mode return to the initial state over the inverse of the intrinsic mechanical decay rate $\gamma$. Or, if the quality factor of the mechanical mode $Q$ is too high for this too be practical, we assume that the mechanical mode can be brought back to the initial state over $10^3$ mechanical periods, e.g. via active-feedback. Furthermore, in order to obtain sufficient experimental statistical data, many runs of the experiment will need to be completed. Supposing that one thousand runs of the experiment will be sufficient, it will take a total time of approximately
\begin{equation} \label{timetaken}
T_{\mathrm{tot}}=10^{3}(2\pi{N}/\omega+T_{\mathrm{r}})/P_{N}. 
\end{equation}
Note that this expression can be used for both of the optical inputs considered in this work by appropriate choice of $P_{N}$. 

Furthermore, it is desirable that the decoherence time of the mechanics $1/\Gamma$ is much greater than the time taken to perform a run of the experiment $2\pi{N}/\omega$ in order to limit the effects of decoherence in the state preparation. Here $\Gamma=(2\bar{n}_{\mathrm{b}}+1)\gamma$ and $\bar{n}_{\mathrm{b}}$ is the occupation of the thermal bath, which is not necessarily in thermal equilibrium with the mechanical mode. This leads to the condition $Q/(2\bar{n}_{\mathrm{b}}+1)\gg2\pi{N}$, which is easily met in most ultracold atom implementations and solid-state optomechanical systems operating at cryogenic temperatures. Proposed complete parameter sets will be discussed in Section \ref{results}.

It is constructive to compare the heralding probabilities of the protocol introduced here with the multi-photon counting scheme introduced in Ref.~\cite{Ringbauer2016}. There, mechanical state generation is achieved via the interaction with a coherent state of light $\ket{\sqrt{2}\alpha}$, followed by a projection of the optical mode onto a NOON state. If $N_{\mathrm{p}}$ is the size of multi-port implemented in this scheme, then the heralding probability for state generation is given by $P_{N_{\mathrm{p}}}=2N_{\mathrm{p}}^{-N_{\mathrm{p}}}\mathrm{e}^{-2|\alpha|^2}|\alpha|^{2N_{\mathrm{p}}}\Big\{1-(-1)^{N_{\mathrm{p}}}\exp[-N_{\mathrm{p}}^2\mu^2(1+2\bar{n})/4]\cos(N_{\mathrm{p}}\phi)\Big\}$. By choosing $\phi=\pi$, we may therefore rewrite this heralding probability---along with the two heralding probabilities from Eq.~(\ref{heralding})---as a scaling $S$ multiplied by $\Big\{1-\exp[-N'^2\mu^2(1+2\bar{n})/4]\Big\}$, where $N'=N$ or $N_{\mathrm{p}}$. Optimizing these scalings over coherent amplitudes, and assuming that $\eta=1$, leads to
\numparts
\begin{eqnarray}
      S_{\mathrm{(i)}}=2^{1-2N}\mathrm{e}^{-N} \label{scalingcoherent}\\
      S_{\mathrm{(ii)}}=2^{1-N} \label{scalingphoton} \\
      S_{\mathrm{(iii)}}=2^{1-N_{\mathrm{p}}}\mathrm{e}^{-N_{\mathrm{p}}} \label{scalingfringes},     
\end{eqnarray}
\endnumparts
where (i) and (ii) refer to the scalings for ${\ket{\psi}}={\ket{\sqrt{2}\alpha}}$ and $\ket{\psi}=\ket{1}$ in our multistep scheme, respectively, while (iii) is the optimal scaling from Ref.~\cite{Ringbauer2016}.  Comparison of $S_{\mathrm{(i)}}$ with $S_{\mathrm{(iii)}}$ indicates that the heralding probability scales more favourably with multi-port size than with step number for the case of a coherent input ${\ket{\sqrt{2}\alpha}}$. However, Eq.~(\ref{scalingphoton}) shows that the multistep protocol with a single photon input gives the best heralding probability.

With the above probabilities in mind, we would like to further note that the multi-photon counting scheme in Ref.~\cite{Ringbauer2016} and the multistep protocol we introduce here may be combined. In such a scheme, the momentum transfer to the mechanics per step would increase to $N_{\mathrm{p}}\mu$, where $N_{\mathrm{p}}\geq2$, and thus coherent states would be required as the optical input states.

\subsection{Model for decoherence}
The fidelity between the final state and the desired output state of the protocol will be limited by interactions with the thermal environment. Here, we model the decoherence of the mechanical mode through the semigroup mapping~\cite{Musslimani1995,Khosla2018}
\begin{equation} \label{semigroup}
\rho_{\mathrm{out}}=(\pi\bar{n}_{\mathrm{th}})^{-1}\int\mathrm{d}^{2}\beta\mathrm{e}^{-|\beta|^{2}/\bar{n}_{\mathrm{th}}}D(\beta)\rho_{\mathrm{in}}D^{\dagger}(\beta),
\end{equation}
which describes an admixture of thermal phonons to the input state $\rho_{\mathrm{in}}$. In this thermalization process $\bar{n}_{\mathrm{th}}$ can be thought of as the mean number of phonons added to the mechanics by the thermal environment between each step. By expanding Eq.~(\ref{semigroup}) using $\bar{n}_{\mathrm{th}}\rightarrow\bar{n}_{\mathrm{th}}+\delta\bar{n}_{\mathrm{th}}$, to first order in $\delta\bar{n}_{\mathrm{th}}$, we see that the thermalization process is equivalent to the standard single-phonon master equation that describes the evolution of a state in contact with a high temperature bath. In other words, in this limit, the single-phonon master equation may be integrated over the mechanical period to obtain Eq.~(\ref{semigroup}), where $\bar{n}_{\mathrm{th}}$ is then the mean number of thermal phonons added to the state during this time. Mathematically, the requirement for a high temperature bath is $\bar{n}_{\mathrm{b}}\gg1$, which remains an appropriate limit for many mechanical quantum systems even at cryogenic temperatures.
 
As in Section \ref{basicprotocol} we assume that the mechanical mode is initially in the thermal state $\rho_{\bar{n}}$ and we apply the measurement operator $\Upsilon_{0,1}^{(j)}$ at each step. Eq.~(\ref{semigroup}) is now used to model the decoherence during the mechanical period between each step. This may be computed analytically, and the Wigner distribution after $N$ iterations of this step-thermalization process is

\begin{eqnarray} \label{Wnr}
\fl W_{\bar{n}_{\mathrm{th}}}(X,P)=\mathcal{N}\sum_{l_{1},m_{1},\ldots,l_{N},m_{N}=0}^{1}\Bigg\{\exp[{\rmi}\sum_{i=1}^{N}(l_{i}-m_{i})(\mu{X}-\phi_{i}-\pi)-\frac{1}{2}\bar{n}_{\mathrm{th}}\sum_{i=1}^{N}\xi_{i}^{2}] \nonumber \\
\exp\Big[-\frac{(X+\rmi\bar{n}_{\mathrm{th}}\sum_{i=1}^{N}\xi_{i})^2}{1+2\bar{n}+2N\bar{n}_{\mathrm{th}}}-\frac{(P-\frac{1}{2}\sum_{i=1}^{N}(l_{i}+m_{i})\mu)^2}{1+2\bar{n}+2N\bar{n}_{\mathrm{th}}}\Big]\Bigg\}
\end{eqnarray}
where $\xi_{i}=\sum_{j=1}^{i}(l_{j}-m_{j})\mu$ and $\mathcal{N}$ ensures normalization. For $\bar{n}_{\mathrm{th}}=0$, this Wigner distribution returns to Eq.~(\ref{Wigscseq}) so we may identify the $\bar{n}_{\mathrm{th}}$ dependent terms in Eq.~(\ref{Wnr}) as a damping and shifting of the interference features of the Wigner distribution. This decoherence effect leads to an imperfect cancellation of the $N$ population terms in the mechanical cat state. Given that the decoherence timescale $1/\Gamma$ is the time taken for the thermal environment to introduce half a phonon to the mechanical mode, the number of phonons added between each step may be approximated as 
\begin{equation}\label{nr}
\bar{n}_{\mathrm{th}}\approx(2\pi/\omega_{\mathrm{M}})\Gamma/2=\pi(2\bar{n}_{\mathrm{b}}+1)/Q.
\end{equation}

\section{Discussion and results}
In this section we discuss the non-classicality and macroscopicity of the states produced by this protocol. We then demonstrate that significant non-classicality and macroscopicity for realistic experimental scenarios can be generated. The non-classicality measures we investigate are based on the negativity of the Wigner distribution, while the macroscopicity measures we study are the phase-space measure $\mathcal{I}$~\cite{Lee2011} and a measure based on the quantum Fisher information~\cite{Braunstein1994}. Definitions and descriptions of these non-classicality and macroscopicity measures will be given below.
\subsection{Non-classicality}\label{non-classicality}
The complementarity between the canonical quadrature operators is manifest in the mathematical properties of the Wigner quasiprobability distribution. Specifically, Wigner distributions may become negative, in stark contrast to classical joint probability distributions, and hence negativity in the Wigner distribution is a signature of non-classicality. Therefore, we present two measures of non-classicality based on the Wigner distribution---its minimum value ($\mathrm{min}~W$) and its total negative volume $\delta$~\cite{Kenfack2004}.

The absolute minimum value any properly normalized Wigner distribution may take is $-1/\pi$. When the multistep protocol is applied to an initial thermal state, and decoherence is ignored between steps, the final mechanical state is given by Eq.~(\ref{Wigscseq}). In this case, the minimum value of $W_{\mathrm{SCS}}$ occurs at $X=0$, $P=N\mu/2$ and is given by 
\begin{equation} 
\mathrm{min}~W_{\mathrm{SCS}}=-\frac{1}{\pi(1+2\bar{n})}\frac{\{1-\exp[-N^2\mu^2/4(1+2\bar{n})]\}}{\{1-\exp[-N^2\mu^2(1+2\bar{n})/4]\}}.
\end{equation}
For this state, in the limit of large momentum separation, the minimum of the Wigner distribution becomes $-{1}/{\pi(1+2\bar{n})}$, and therefore the states produced in our multistep protocol approach the absolute minimum value for $\mathrm{min}~W$ in the limit $\bar{n}\rightarrow0$.

The total negative volume of the Wigner distribution $\delta$~\cite{Kenfack2004} is calculated using
\begin{equation}
\delta=\frac{1}{2}\bigg(\int\int|W(X,P)|\mathrm{d}X\mathrm{d}P-1\bigg),
\end{equation}
which is strictly greater than or equal to zero. Furthermore, in \ref{appendixmax} we show that for SCSs this measure is always less than $1/\pi$. This upper bound is reached for $N\mu\gg1$ even in the case where the mechanical mode supports an initial non-zero mean thermal occupation. However, the introduction of mechanical decoherence ($\bar{n}_{\mathrm{th}}\neq0$) between steps causes the value of $\delta$ to reduce and drop below this maximal value, as will be discussed in the results section below.

\subsection{Macroscopicity}\label{macroscopicity}
The phase-space measure of macroscopicity for a single-mode Wigner distribution introduced by Lee and Jeong in Ref.~\cite{Lee2011} is defined as
\begin{equation}
\mathcal{I}=-\frac{\pi}{2}\int_{-\infty}^{+\infty}\int_{-\infty}^{+\infty}W(X,P)(\nabla^2+2)W(X,P)\mathrm{d}X\mathrm{d}P,
\end{equation}
where $\nabla^2$ is the Laplacian in phase space. $\mathcal{I}$ quantifies the size of the superposition by measuring the degree to which the sharp features, caused by interference effects, extend in phase space. Lee and Jeong show that pure SCSs belong to a class of states that saturate the upper bound that can be reached by $\mathcal{I}$, namely the expectation value of the number operator.

As a second method to quantify macroscopicity we use a measure based on quantum Fisher information (QFI) $\mathcal{F}$. QFI is typically used as a tool in parameter estimation~\cite{Demkowicz-Dobrzanski2014}, however, more recently QFI has been used as a genuine measure of macroscopic quantum effects in spin ensembles~\cite{Frowis2012} and in photonic systems~\cite{Oudot2015}. The macroscopicity of a state $\rho$ can be quantified by its sensitivity to translations in phase space. First note that the translated state is given by $\rho_{\tau}=\mathrm{e}^{-\rmi{\tau}X_{\theta}}\rho\mathrm{e}^{\rmi\tau{X_{\theta}}}$, with $X_{\theta}=(a\mathrm{e}^{-\rmi\theta}+a^{\dagger}\mathrm{e}^{\rmi\theta})/\sqrt{2}$. The QFI, $\mathcal{F}_{\tau}$, of these translated states is then calculated for each $\theta$, and finally a maximization over all $\theta$ is performed. Therefore the authors of Ref.~\cite{Oudot2015} propose that for a single mode of a bosonic system the `effective size' of the state $\rho$, and the appropriate measure of macroscopicity, is given by $N_{\mathrm{eff}}(\rho)=\frac{1}{2}\max\limits_{\theta}\mathcal{F}_{\tau}$. Encouragingly, QFI has been shown to satisfy a set of conditions for measures of macroscopic coherence in Ref.~\cite{Yadin2016}. Therefore we will now present a measure based on QFI as an alternative measure of macroscopicity. We will also present $\mathcal{I}$ in the results section to allow for comparison with other macroscopic state generation schemes, e.g. Ref.~\cite{Hoff2016}. 

The measure $N_{\mathrm{eff}}(\rho)$ is difficult to calculate for mixed states, as diagonalization of the density matrix---or, equivalently, a maximization over all POVMS---is required to obtain the QFI. Whereas for pure states no such diagonlization or maximization is required. In particular, for the pure SCS, with $\bar{n}=\bar{n}_{\mathrm{th}}=0$, $\mathcal{F}_{\tau}=2+N^2\mu^2/[1-\exp(-N^2\mu^2/4)]-N^2\mu^2\cos^2\theta$, which is maximimum at $\theta=\pi/2$. To alleviate the difficulty of maximizing over all possible POVMS when the state is mixed, we restrict ourselves to the set of POVMs given by $E(X_{\lambda})=\ket{X_{\lambda}}\bra{X_{\lambda}}$, where $0\leq\lambda<\pi$, which corresponds to quadrature measurements at an angle $\lambda$ from the position axis. In this work we therefore use a macroscopicity measure $\mathcal{M}$ that satisfies $\mathcal{M}\leq\mathcal{N}_{\mathrm{eff}}(\rho)$. As detailed in \ref{macroappendix}, the above restriction on the set of all POVMS leads to a simple expression for the macroscopicity measure $\mathcal{M}$, given by $\mathcal{M}=\frac{1}{2}\max\limits_{\lambda}F_{X_{\lambda}}$. Here, $F_{X_{\lambda}}$ is the classical Fisher information (CFI) with respect to the quadrature $X_{\lambda}$. Furthermore, the states which are produced in our multistep protocol contain interference terms that oscillate along the X axis of phase space, and hence the maximization is achieved at $\lambda=0$. Further note that the CFI of the SCS, with $\bar{n}\neq0$ and $\bar{n}_{\mathrm{th}}=0$, is given by $F_{X}=2/(2\bar{n}+1)+N^2\mu^2/(1-\exp[-N^2\mu^2(2\bar{n}+1)/4])$. So, in the case of a pure SCS, $\bar{n}=\bar{n}_{\mathrm{th}}=0$, it is clear that a quadrature measurement along $X$ is in fact the optimal POVM measurement, which provides further motivation for restricting the class of POVMs to quadrature measurements. 

\subsection{Results}\label{results}

In this multistep protocol, the non-classicality and macroscopicity of the final state depends on $N\mu$ and the initial thermal occupation of the mechanical mode. As an example, consider the ideal case $\bar{n}=\bar{n}_{\mathrm{th}}=0$, where the RMS width of the initial Gaussian state is $\sqrt{\braket{X^2}}=1/\sqrt{2}$. For $N\mu<1/\sqrt{2}$, the final state of the mechanical mode resembles a single-phonon Fock state. This can be seen by taking the limit $N\mu\rightarrow0$ in Eq.~(\ref{rhoout2}), which gives $[D(\rmi{N}\mu/\sqrt{2})-\mathds{1}]\ket{0}\approx\rmi{N}\mu{X}\ket{0}\propto\ket{1}$. Varying the value of $N\mu$ in the range $N\mu<1/\sqrt{2}$ has very little effect on the non-classicality and macroscopicity measures of these states. For example, when $N\mu$ is increased from $10^{-4}$ to $0.1$, $\delta$ remains approximately $0.2131$, this equalling the value of $\delta$ for a single-phonon Fock state $\delta_{\ket{1}}=2\mathrm{e}^{-0.5}-1\simeq0.2131$. The macroscopicity measure $\mathcal{M}$ also remains clamped in this interval at $3$---this being equal to the value of $\mathcal{M}$ for the state $\ket{1}$. When $N\mu$ is increased above the width of the initial Gaussian state the state becomes more macroscopic. For example, when $N\mu={2}$, the non-classicality measure $\delta$ remains saturated at the value of the single-phonon Fock state, whereas $\mathcal{M}$ has increased rapidly to $4.16$. The phase-space distribution now resembles a `kitten' state---a SCS with a small separation between population terms. As $N\mu$ is increased even further, such that $N\mu>4/\sqrt{2}$, the non-classicality and macroscopicity measures increase more dramatically. At $N\mu=4$, $\delta$ reaches $0.2462$ and $\mathcal{M}$ increases to $9.15$. When $N\mu>4/\sqrt{2}$, the separation of the two Gaussian population terms is greater than four times the RMS width of the ground state, such that the diameter of the interference fringes---given by twice the RMS width of the ground state--- does not overlap with the width of the population components. Moreover, when $N\mu>4/\sqrt{2}$ the mechanical mode is a SCS as the overlap between the two Gaussian population terms is negligible (less than $1.83\%$).

\begin{figure}[H]
\begin{framed} \centering
\includegraphics[scale=0.22,angle=270,trim=0 1.25cm 0 0]{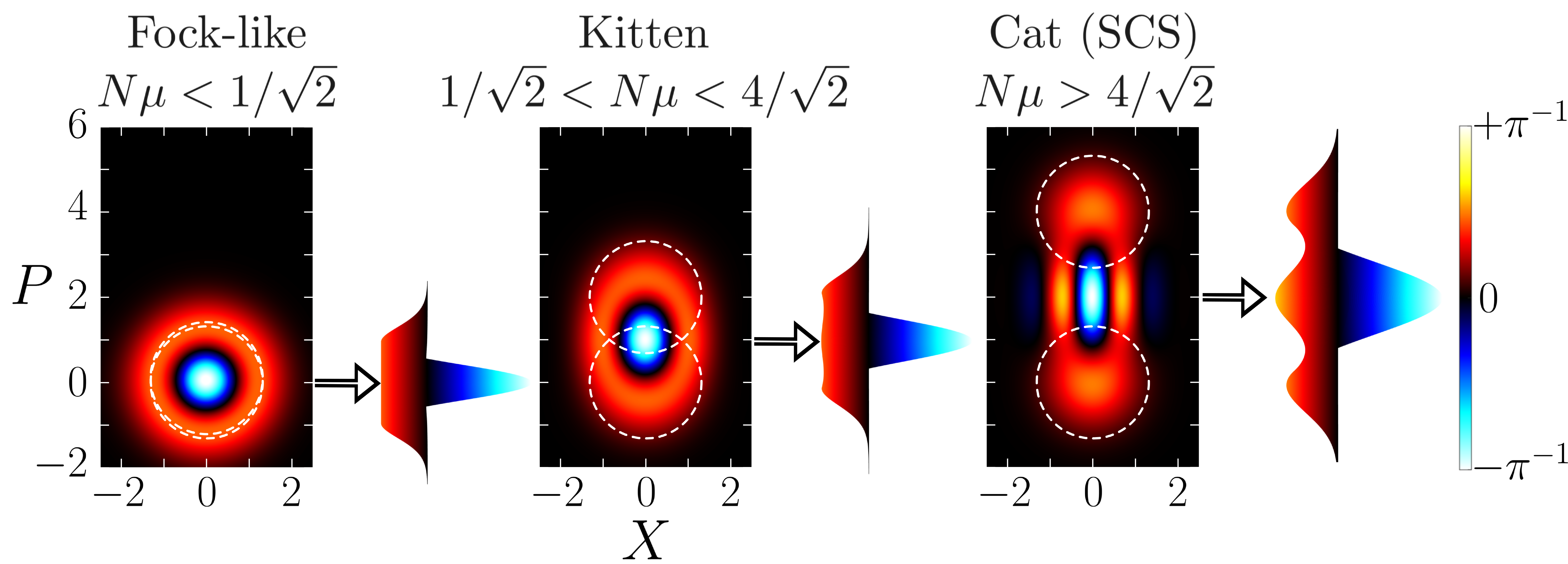} 
{\color{black}\caption{\small{{\textbf{Wigner distributions of the single-phonon Fock-like state, kitten state, and Schr\"{o}dinger cat state that can be generated by our scheme}}. This figure shows how the ratio of $N\mu$ and the initial width of the mechanical mode determines the nature of the final state produced in the multistep protocol. Here, $\bar{n}=0$, $\bar{n}_{\mathrm{th}}=0$ and $N\mu$ increases from left to right as $0.1$, $2$ and $4$. White-dashed circles of radius $\sqrt{2}$---twice the RMS width of the ground state---are placed around each of the Gaussian population terms. When $N\mu>4/\sqrt{2}$ these circles do not overlap and the mechanical mode is in a cat state. Side-views of the Wigner distributions parallel to the position axis are also plotted to highlight the difference between the three different regimes of states.}}
\label{regimes}}
\end{framed}
\end{figure}

In the impure case, initial thermal occupation, or thermal decoherence, leads to a smearing of the population terms and a reduction in the quantum interference fringe visibility. The decoherence process modelled by Eq.~(\ref{semigroup}) reduces the non-classicality and macroscopicity of the state in a non-trivial way as the state grows, such that at a certain step number the non-classicality and macroscopicity of the may begin to decrease as the multistep protocol proceeds. This is discussed in more detail below and computed numerically. Fig.~\ref{regimes} illustrates the three different regimes of states generated in the protocol depending upon the value of $N\mu$. In particular, the side-view of this figure shows how the two peaks from the Gaussian population terms in the phase-space distribution begin to become resolvable when $N\mu$ increases past $1/\sqrt{2}$, before becoming fully distinct when $N\mu>4/\sqrt{2}$. 

Table \ref{table1} lists a range of parameter sets taken from recent experiments that implement ultracold atom or solid-state realizations of optomechanical systems. In order to expose the mechanical quantum features, which lie beneath any thermal fluctuations, the mechanical system must be cooled close to its ground state. Therefore, in this table for the solid state devices it is assumed that the mechanical resonator is pre-cooled to $\bar{n}=0.1$ and the bath temperature is $100~\mathrm{mK}$. This can be done using a combination of cryogenics, back-action cooling and active-feedback cooling methods~\cite{RMP2014}. The table also includes a range of proposed parameter sets that enable improved performance with our protocol. We consider the case where the optical input state is a single-photon pulse, the step number is $N=3$, and one thousand runs of the multistep protocol have been completed at an optical efficiency of $90\%$, such that $\eta=0.9$. Therefore, the column headed $T_{\mathrm{tot}}$ indicates the approximate total time taken to complete an experiment where one thousand statistical data points are obtained. If a weak coherent pulse is chosen as the input state then the protocol takes approximately one hundred times longer when $N=3$. The final four columns show the values of the non-classicality and macroscopicity measures calculated using Eq.~(\ref{Wnr}) and the definition of these measures from Section \ref{non-classicality} and \ref{macroscopicity}. 

We would like to highlight that the generation of true macroscopic SCS is currently achievable using ultracold implementations of optomechanical devices over realistic timescales. The proposed parameter set (i) is currently realizable and serves as a suggestion for slight improvements that can be made to individual parameters for an even better performance of the scheme. The solid state systems in Ref.~\cite{Wilson2015} and~\cite{Leijssen2017}, as well as proposal (ii), operate in the regime where $N\mu<1/\sqrt{2}$. The protocol is successful in generating macroscopic states for the parameters from Ref.~\cite{Wilson2015} and (ii), as macroscopicity values much larger than that of the initial thermal state are generated. For reference, the macroscopicity of a thermal state is given by $\mathcal{M}=1/(2\bar{n}+1)$, which equals 0.83 for $\bar{n}=0.1$. However, in this regime true SCSs cannot be generated as we require $N\mu>4/\sqrt{2}$. Given the current rate of experimental advances, it is encouraging that $\mu$ need only be increased by a factor of one hundred to allow for the generation of SCSs in solid-state devices when $N=3$. Importantly, this requirement on the improvement of $\mu$ is relaxed even further when $N$ is increased. Proposal (iii) illustrates the success of the protocol in preparing macroscopic SCSs when this requirement on $N\mu>4/\sqrt{2}$ is met. 

The parameter set from Ref.~\cite{Leijssen2017} generates little non-classicality and macroscopicity despite the relatively large value of $\mu$. This is due to the rapid thermalization of the mechanical mode in between steps, which is a consequence of the low quality factor. The performance of the protocol rapidly diminishes as $\bar{n}_{\mathrm{th}}$ is increased in magnitude past $10^{-2}$, and note that the value of $\bar{n}_{\mathrm{th}}$ depends upon the quality factor and the temperature of the thermal bath.

Values for the mean number of phonons added to the mechanical mode were calculated using Eq.~(\ref{nr}). The values of the measures of non-classicality and macroscopicity reached in this multistep protocol are sensitive to the value of $\bar{n}_{\mathrm{th}}$, in fact, for a given value of $\mu$ and $\bar{n}_{\mathrm{th}}$, there may be a point where increasing the step number past a certain value leads to a reduction in the non-classicality and macroscopicity of the state. The plots below illustrate this point and demonstrate that for a given parameter set an optimal step number can be chosen by considering the total time required and calculating the measures from Eq.~(\ref{Wnr}). We now turn our attention to study the non-classicality and macroscopicity of the mechanical states in more detail.

 \begin{table}[H]
 \small
\centering
 \caption{\small{{\textbf{Present-day experimental and proposed parameter sets}}. Here, one thousand runs of the multistep protocol have been considered using single photons as the optical input states. Each run consists of $N=3$ steps with $10\%$ optical loss at each step. All ultracold atom setups are assumed to be operating at $\bar{n}=0$ and at a low bath occupancy, while we assume that the solid-state setups are operating at $\bar{n}=0.1$ in a thermal bath which has been cooled to $100~\mathrm{mK}$. Parameter set (i) is a currently realizable proposal for generating macroscopic SCSs in ultracold atoms, while parameter sets (ii), (iii) are proposals for generating macroscopic states in near-future solid-state systems.}}
 \label{table1}
\setlength\tabcolsep{3pt}
 \begin{tabular}{ c|c|c|c|c|c|c|c|c|c} 
 \hline\hline
Ref. & $\mu$ & $\omega/2\pi\mathrm{(Hz)}$ & $Q$ &$\bar{n}_{\mathrm{th}}$ & $T_{\mathrm{tot}}\mathrm{(s)}$ & $\mathrm{min}~W$ & $\delta$ & $\mathcal{I}$ & $\mathcal{M}$ \\ 
 \hline
 \multicolumn{10}{c}{Ultracold atom systems}\\
 \hline
\cite{Brennecke2008} & $17.8$ & $37\times10^{3}$ & $581$ &$5.41\times10^{-3}$& $14.1$ &$-0.047$ &$0.115$ &$1.71$ &$11.6$ \\ 
\cite{Purdy2010} &  $15.4$ & $50\times10^3$ & $314$ &$7.14\times10^{-3}$& $5.82$  &$-0.046$ &$0.116$ &$1.28$ &$9.06$  \\ 
(i) & $15.0$ & $50\times10^3$ & $785$  &$4.00\times10^{-3}$& $14.0$ & $-0.049$ & $0.142$ & $2.05$ & $13.7$  \\
 \hline
 \multicolumn{10}{c}{Solid-state systems}\\
 \hline
\cite{Wilson2015} & $9.64\times10^{-5}$ & $4.30\times10^6$ & $7.54\times10^5$  &$4.05\times10^{-3}$& $5.10\times10^7$ & $-0.172$ & $0.117$ & $0.323$ & $1.91$ \\
\cite{Leijssen2017} & $8.44\times10^{-3}$ & $3.74\times10^6$ & $3.74\times10^4$  &$9.40\times10^{-2}$& $7.65\times10^3$ & $-0.023$ & $0.010$ & $-0.032$ & $0.512$ \\
(ii) & $0.10$ & $1.00\times10^6$ & $6.28\times10^6$  &$2.09\times10^{-3}$& $207$ & $-0.178$ & $0.122$ & $0.351$ & $2.09$ \\
(iii) & $1.00$ & $1.00\times10^6$ & $6.28\times10^6$  &$2.09\times10^{-3}$& $5.90$ & $-0.230$ & $0.165$ & $0.886$ & $4.39$ \\
\hline\hline
\end{tabular}
\end{table}

\begin{figure}[H]
\begin{framed} \centering
\includegraphics[scale=0.23, angle=270]{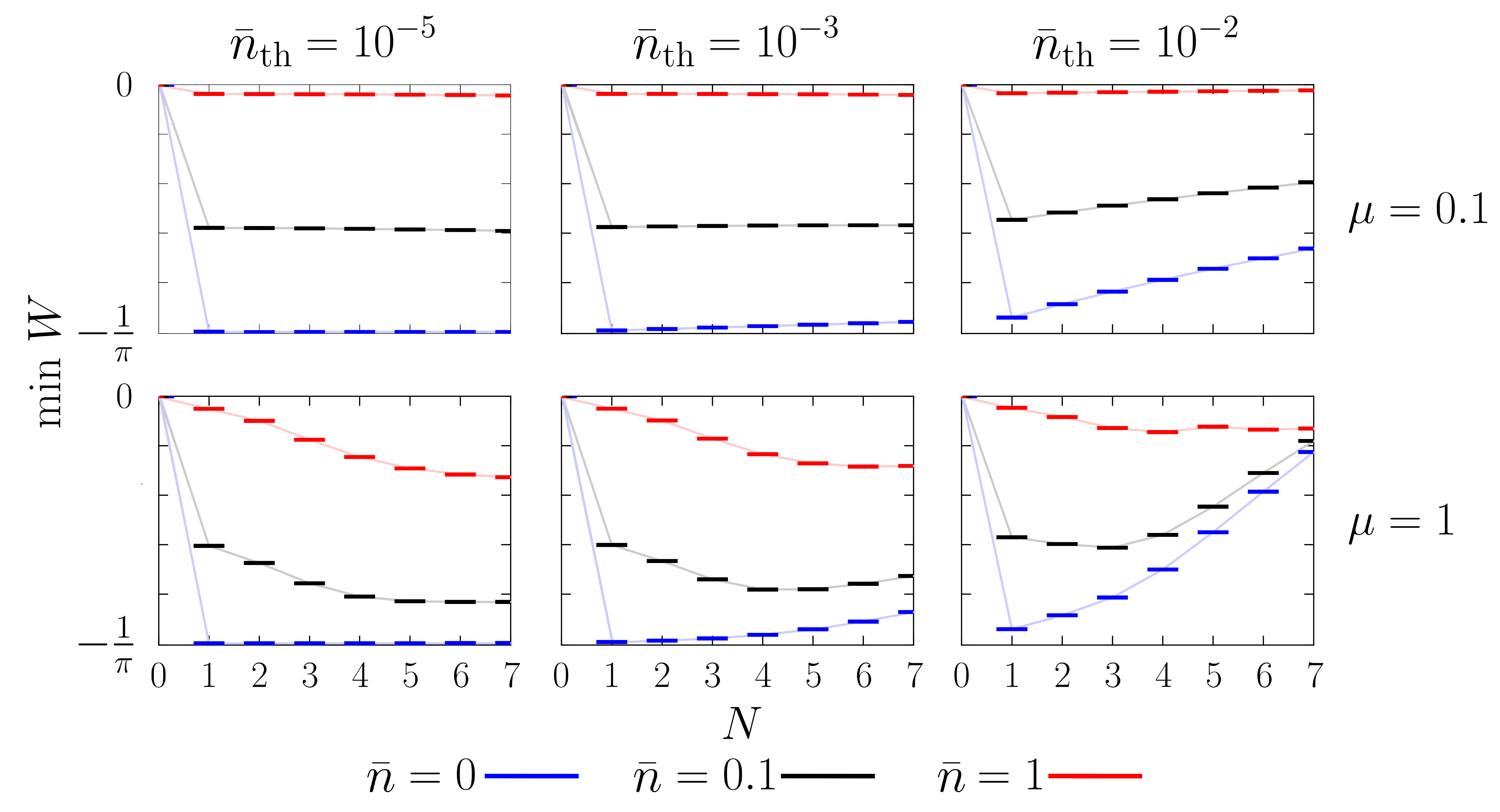} 
\caption{\small{{\textbf{Minimum of the Wigner distribution $\bm{\mathrm{min}~W}$ plotted as a function of step number $\bm{N}$ for a range of $\bm{\mu}$, $\bm{\bar{n}_{\mathrm{th}}}$, and $\bm{\bar{n}}$}}. In the ideal case, $\bar{n}=\bar{n}_{\mathrm{th}}=0$, this non-classicality measure saturates at the minimum possible value of $-1/\pi$ irrespective of the value of $\mu$. Although the Wigner distribution has greater negativity for $\mu=1$ and for small decoherence ($\bar{n}_{\mathrm{th}}=10^{-5}$), the lower value of $\mu$ produces a state with a $\mathrm{min}~W$ more resilient to the thermal decoherence as $N$ increases.}}
\label{plots1}
\end{framed}
\end{figure}

\begin{figure} 
\begin{framed}\centering
\includegraphics[scale=0.23, angle=270]{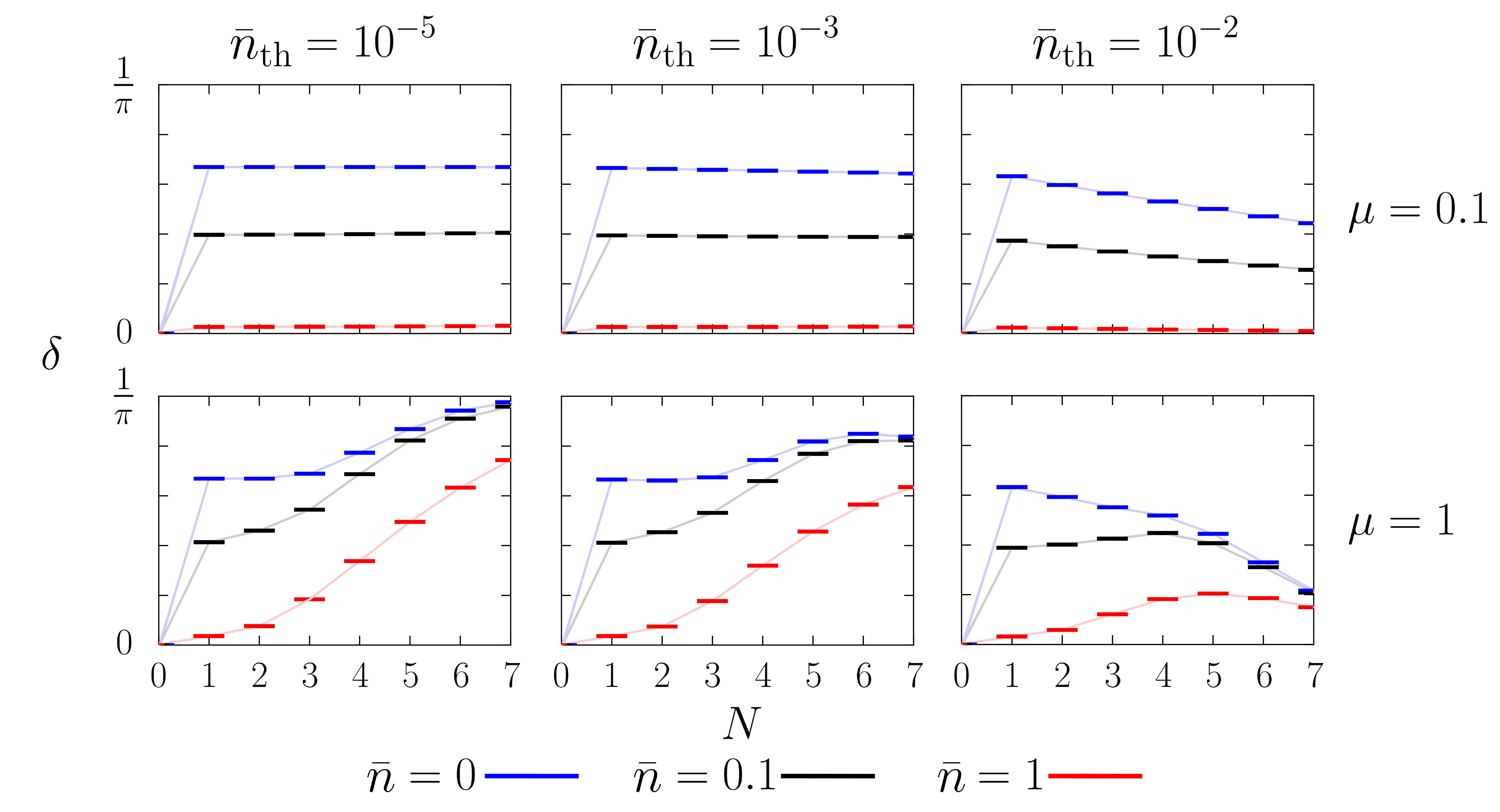} 
\caption{\small{{\textbf{Total negative volume in phase space $\bm{\delta}$ plotted as a function of step number $\bm{N}$ for a range of $\bm{\mu}$, $\bm{\bar{n}_{\mathrm{th}}}$, and $\bm{\bar{n}}$}}. States produced in the large $N\mu$ regime with $\bar{n}_{\mathrm{th}}=0$ saturate the maximum value of this non-classicality measure for a SCS state, $\delta_{\mathrm{max}}=1/\pi$, independent of the value of $\bar{n}$. The plots also show that for $\bar{n}=0$, $\bar{n}_{\mathrm{th}}\approx0$, and $N\mu<4/\sqrt(2)$ the value of $\delta$ is equal to that of the single-phonon Fock state as discussed above in the main text. The states with lower phase-space separation are better at retaining their $\delta$ values as $N$ increases when decoherence effects are considered.}}
\label{plots2}
\end{framed}
\end{figure}

\begin{figure}
\begin{framed}\centering
\includegraphics[scale=0.23,angle =270]{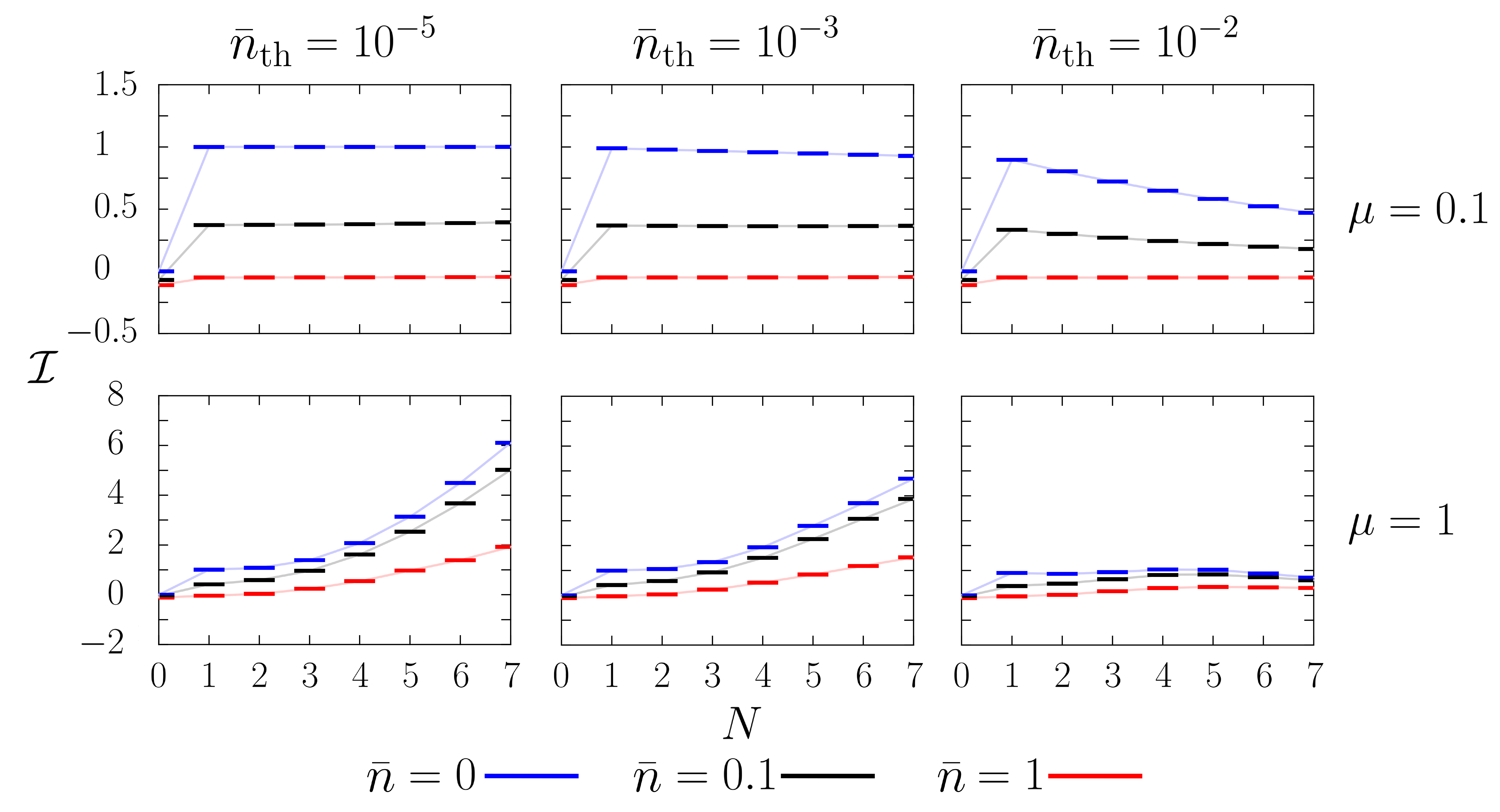} 
\caption{\small{{\textbf{The phase-space macroscopicity measure $\bm{\mathcal{I}}$ plotted as a function of step number $\bm{N}$ for a range of $\bm{\mu}$, $\bm{\bar{n}_{\mathrm{th}}}$, and $\bm{\bar{n}}$.}} With the set of parameters denoted in the figure, the plots demonstrate the regime where the multistep protocol is successful in growing macroscopicity---as measured by $\mathcal{I}$---with step number $N$. The plots also identify the regime where $\mathcal{I}$ decreases with step number when decoherence effects begin to dominate.}}
\label{plots3}
\end{framed}
\end{figure}

\begin{figure}[H]
\begin{framed}\centering
\includegraphics[scale=0.23, angle=270]{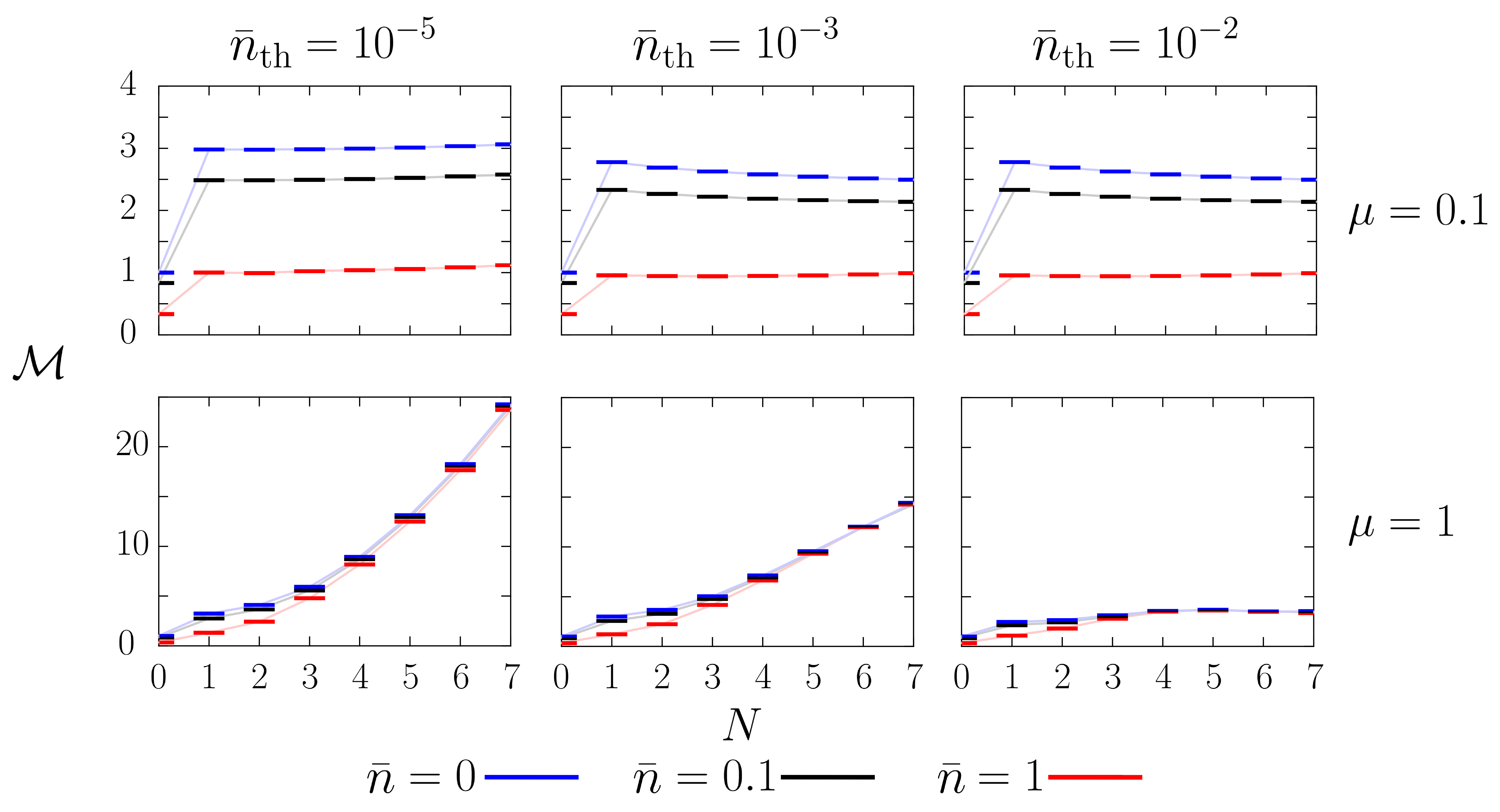} 
\caption{\small{{\textbf{The quantum Fisher information based macroscopicity measure $\bm{\mathcal{M}}$ plotted as a function of step number $\bm{N}$ for a range of $\bm{\mu}$, $\bm{\bar{n}_{\mathrm{th}}}$, and $\bm{\bar{n}}$.}} $\mathcal{M}$ scales similarly to $\mathcal{I}$, reflecting the similarity of the two measures, and the same general results discussed in the caption of Fig.~\ref{plots3} apply here too. However, as compared to Fig.~\ref{plots3}, the increase in macroscopicity with step number $N$ is just visible when $\mu=0.1$ and $\bar{n}_{\mathrm{th}}=10^{-5}$.}}
\label{plots4}
\end{framed}
\end{figure}

The plots in Figs.~\ref{plots1} through \ref{plots4} consider the dependence of the non-classicality and macroscopicity measures on step number for $\bar{n}_{\mathrm{th}}=\{10^{-5},10^{-3},10^{-2}\}$, $\bar{n}=\{0,0.1,1\}$, and $\mu=\{0.1,1\}$. As expected, the introduction of decoherence between steps filters out the high-frequency contributions to the Wigner distribution, which leads to a decrease in the non-classicality and macroscopicity values. We consider step numbers ranging from zero through to seven and note that the case where $\bar{n}_{\mathrm{th}}=10^{-5}$ performs as well as the ideal case, with $\bar{n}_{\mathrm{th}}=0$, to within five percent by the end of the seventh step for all measures. Therefore, if optomechanical devices could be produced with sufficiently high quality factors operating at millikelvin temperatures, such that $\bar{n}_{\mathrm{th}}\approx10^{-5}$, then Eq.~(\ref{Wigscseq}) would serve as an accurate tool for analysis up to at least seven steps. We also consider the cases where $\bar{n}_{\mathrm{th}}=10^{-3}$, which was considered in Table \ref{table1}, and $\bar{n}_{\mathrm{th}}=10^{-2}$, which represents a situation where thermal interactions are much stronger. Furthermore, we assume that either single photons or weak coherent states, satisfying $(1-\eta)|\alpha|^2\ll1$, are used as input states such that the presence of optical loss has no effect on the phase distribution of the final mechanical state.

The non-classicality measures plotted in Figs.~\ref{plots1} and \ref{plots2} are very sensitive to initial thermal occupation, which reduces the prominence of any non-classical features of the final mechanical state. The amount of thermal occupation, and therefore the reduction in `phase-space sharpness', is further increased by the decoherence process modelled in between steps of the protocol. Generally speaking, Figs.~\ref{plots1} and \ref{plots2} show that states with the highest values of $N\mu$, and therefore the sharpest phase-space features, are more susceptible to thermal decoherence effects. This is the conventional quantum-to-classical transition in action. Note that, even after seven steps of our multistep protocol the states with $\mu=1$ experiencing heavy thermalization, $\bar{n}_{\mathrm{th}}=10^{-2}$, still retain significant non-classicality---approximately one-fifth of the maximal values for each non-classicality measure. 

The macroscopicity measures are plotted in Figs.~\ref{plots3} and \ref{plots4} and by comparing these plots we see that these two measures scale in a similar way with $N$. This is because both $\mathcal{I}$ and $\mathcal{M}$ measure macroscopicity via the degree to which sharp features extend in phase space. $\mathcal{I}$ measures this sharpness by averaging over the whole of phase space, while $\mathcal{M}$ selects an optimal direction in phase space. The multistep protocol discussed above is deemed successful if it can grow mechanical superposition states towards macroscopic values even in the presence of thermal occupation and decoherence. For the chosen parameter range, Figs.~\ref{plots3} and \ref{plots4} show that this goal is possible for $\mu=0.1$ only for small $\bar{n}_{\mathrm{th}}$. The plots for $\mu=0.1$ and $\bar{n}_{\mathrm{th}}=10^{-3},10^{-2}$ show a decrease in $\mathcal{I}$ and $\mathcal{M}$ with step number. While the corresponding plot at $\bar{n}_{\mathrm{th}}=10^{-5}$ produces an increase in these measures with step number---this increase is slight and not visible in Fig.~\ref{plots3}. But perhaps more interestingly, for $\mu=1$, the macroscopicity of the decohered states increases as the state grows for $N\leq5$ even in the case of heavy thermalization, $\bar{n}_{\mathrm{th}}=10^{-2}$. This demonstrates that our protocol is capable of generating significant macroscopicity even when decoherence and initial thermal occupation of the mechanical mode are accounted for. It must however be noted that past $N=5$ the protocol leads to a decrease in macroscopicity, meaning that $5$ steps is the optimal step number for $\mu=1$ and $\bar{n}_{\mathrm{th}}=10^{-2}$. For a given coupling strength and decoherence rate this result indicates that by analysing Eq.~(\ref{Wnr}) an optimal step number can be established. 


\section{Conclusion}

In this work, we have proposed a measurement-based multistep protocol for macroscopic quantum state preparation of mechanical motion via cavity quantum optomechanics. Our protocol not only allows for the increase in non-classicality and macroscopicity with step number, but also relaxes the requirement on the optomechanical coupling strength needed to prepare well-separated mechanical superposition states. We have focused primarily on the preparation of mechanical Schr\"{o}dinger cat states, which consist of a superposition of two distinct momentum states. This is achieved by setting the phase within an optical interferometer, at each step of the $N$-step protocol, to be the $N^{\mathrm{th}}$ root of unity. We would also like to highlight that by varying the phase $\phi_{j}$ introduced at each step, our scheme opens up the possibility of preparing a broad range of non-classical mechanical states.

Our protocol may be implemented with readily-available optical inputs: single photons or weak coherent states. Both inputs provide resilience to optical loss, where a more favourable heralding probability is obtained with single photons. Using parameters from recent experiments we have studied the effect that initial thermal occupation and mechanical decoherence has on the heralded mechanical state. We have shown that our scheme may be used to grow macroscopic Schr\"{o}dinger cat states in present-day ultracold atom implementations of cavity optomechanical systems. Furthermore, we have also shown that our scheme provides an experimentally feasible route to grow macroscopic mechanical superposition states in solid-states devices using present-day, and expected near-future, experimental parameters.  

This work provides a new path towards the longstanding goal of quantum state generation at a truly macroscopic scale, which would allow for the study of wavefunction collapse and the interface between quantum mechanics and gravity to be explored. We would also like to note that this protocol can be readily adapted to be employed in systems which interact via a spin-phonon coupling, e.g. of the form $H_{\mathrm{int}}=\hbar\lambda\sigma_{z}X$. Prominent examples of such systems include mechanical oscillators with nitrogen-vanancy centres~\cite{Arcizet2011,Yin2013} and superconducting electromechanical platforms~\cite{Lecocq2015,Khosla2018}.

\section*{Acknowledgements}
We would like to thank Rufus Clarke, Kiran E. Khosla, M. S. Kim, Hyukjoon Kwon, Martin Ringbauer, Andrew G. White, and Benjamin Yadin for stimulating discussions. This research was funded by the Engineering and Physical Sciences Research Council (EP/N014995/1).

\vfill

\appendix
\section{Pulsed optomechanical interactions}\label{pulseshapeappendix}
In the multistep pulsed protocol we have introduced, we require operation well outside of the resolved sideband regime so that the mechanical free evolution may be neglected during the interaction, i.e. we require $\kappa\gg\omega$. Furthermore, for any reasonable mechanical oscillator, with a quality factor much greater than one, this means that the damping and noise on the mechanical mode can be also ignored during the light-matter interaction. Under these conditions, for a resonant pulsed drive, in a frame rotating at the cavity frequency, the Langevin equations for the system are
\begin{eqnarray}
\frac{\mathrm{d}a}{\mathrm{d}t}=(\rmi\sqrt{2}Xg_{0}-\kappa)a+\sqrt{2\kappa}a_{\mathrm{in}},  \label{adot} \\
\frac{\mathrm{d}b}{\mathrm{d}t}=\rmi{g}_{0}a^{\dagger}a,
\end{eqnarray}
where we only consider the interaction between a single intracavity field, described by operator $a$, and a single mechanical mode, described by $b$. Here, $a_{\mathrm{in}}$ is the operator for the input field, which obeys the input-output boundary condition $a_{\mathrm{out}}=\sqrt{2\kappa}a-a_{\mathrm{in}}$. As $X$ is approximately constant during the interaction, we integrate Eq.~(\ref{adot}), which leads to

\begin{equation}
a(t)=\sqrt{2\kappa}\mathrm{e}^{(-\kappa+\rmi\sqrt{2}Xg_{0})t}\int_{-\infty}^{t}a_{\mathrm{in}}(t')\mathrm{e}^{(\kappa-\rmi\sqrt{2}Xg_{0})t'}\mathrm{d}t'. \label{a}
\end{equation}
By rewriting the Langevin equations for the mechanical mode in terms of $X$ and $P$ we arrive at

\begin{eqnarray}
\frac{\mathrm{d}X}{\mathrm{d}t}=0, \label{Xdot} \\
\frac{\mathrm{d}P}{\mathrm{d}t}=\sqrt{2}g_{0}a^{\dagger}a. \label{Pdot}
\end{eqnarray}
Eqs.~(\ref{Xdot}) and (\ref{Pdot}) are then readily solved to obtain

\begin{eqnarray}
X'=X,  \\
P'-P=\sqrt{2}g_{0}\int_{-\infty}^{+\infty}{a}^{\dagger}a(t)\mathrm{d}t. \label{momentumexchange}
\end{eqnarray}
Here, the prime indicates the state of the system after the pulsed interaction and the limits of integration are valid if the pulse duration is much shorter than the mechanical period. Immediately, we see that on this short time scale, defined by the temporal envelope of the pulse, the position of the mechanical mode stays constant, while the field imparts a momentum kick to the mechanics.

Now we assume that the input pulse contains a single photon with temporal envelope $f(t)$,
\begin{equation}
\ket{\psi}=\int_{-\infty}^{+\infty}\mathrm{d}t{f}(t)a^{\dagger}_{\mathrm{in}}(t)\ket{0},
\end{equation}
with $\int_{-\infty}^{+\infty}\mathrm{d}t|{f}(t)|^{2}=1$. We may write the momentum exchange during the interaction as $P'-P=\mu{n'}$, which introduces the optomechanical coupling strength. This form of the interaction is evident from the optomechanical unitary $\mathrm{e}^{\rmi\mu{X}{n'}}$ from Section \ref{basicprotocol}, however we are now more careful to distinguish between the number operator integrated over the pulsed-optomechanical interaction $n'$, and the time dependent intracavity field operators $a(t)$. Eq.~(\ref{a}) may be inserted into (\ref{momentumexchange}) to obtain another expression for the momentum exchange
\begin{eqnarray}
P'-P=\sqrt{2}g_{0}\int_{-\infty}^{+\infty}{a}^{\dagger}a(t)\mathrm{d}t=\mu{n'} \nonumber \\
\fl=\sqrt{8}g_{0}\kappa\int_{-\infty}^{+\infty}{\mathrm{d}}t\mathrm{e}^{-2\kappa{t}}\int_{-\infty}^{t}{\mathrm{d}}t'\int_{-\infty}^{t}{\mathrm{d}}t''\mathrm{e}^{(\kappa+\rmi\sqrt{2}Xg_{0})t'}\mathrm{e}^{(\kappa-\rmi\sqrt{2}Xg_{0})t''}a^{\dagger}_{\mathrm{in}}(t')a_{\mathrm{in}}(t'').
\end{eqnarray}
To find an expression for $\mu$ we take the expectation value of the above equation in the total state of the system, which will be a joint state of the external light, the intracavity light, and the mechanical mode. This gives 

\begin{eqnarray}
\braket{P'-P}=\sqrt{2}g_{0}\int_{-\infty}^{+\infty}\braket{a^{\dagger}a(t)}\mathrm{d}t=\mu\braket{{n'}} \nonumber \\
=\sqrt{8}g_{0}\kappa\int_{-\infty}^{+\infty}{\mathrm{d}}t\mathrm{e}^{-2\kappa{t}}\int_{-\infty}^{t}{\mathrm{d}}t'\int_{-\infty}^{t}{\mathrm{d}}t''\mathrm{e}^{\kappa{t'}}\mathrm{e}^{\kappa{t''}}\braket{a^{\dagger}_{\mathrm{in}}(t')a_{\mathrm{in}}(t'')}.
\end{eqnarray}
Where we have used the fact that $\braket{X}=0$. The optomechanical interaction between the intracavity field and the mechanics will generate correlations between the light and mechanical motion, however as the interaction preserves the mechanical position, when the expectation value over the whole system is taken we retain $\braket{X}=0$. The average over the input field operators is given by 
\begin{equation}
\braket{a^{\dagger}_{\mathrm{in}}(t')a_{\mathrm{in}}(t'')}=\bra{\psi}a^{\dagger}_{\mathrm{in}}(t')a_{\mathrm{in}}(t'')\ket{\psi}=f^{*}(t')f(t''),
\end{equation}
and the expectation value of the number operator is $\braket{{n'}}=1$ for a single photon pulse. Finally, the equation for $\mu$ may be written as 

\begin{equation}
\mu=\sqrt{8}g_{0}\kappa\int_{-\infty}^{+\infty}{\mathrm{d}}t\mathrm{e}^{-2\kappa{t}}\int_{-\infty}^{t}{d}t'\int_{-\infty}^{t}{d}t''\mathrm{e}^{\kappa{t'}}\mathrm{e}^{\kappa{t''}}f^{*}(t')f(t''),
\end{equation}
Choosing the pulse shape $f(t)=\sqrt{\kappa}\exp(-\kappa|t|)$, which matches the cavity spectrum, and carrying out the final integral over the entire duration of the pulse leads to $\mu=3g_{0}/\sqrt{2}\kappa$. 

\section{Optical loss}\label{opticalloss}
\subsection{General input state} 
\begin{figure}[h]
\begin{framed}\centering
\includegraphics[scale=0.5, angle=270]{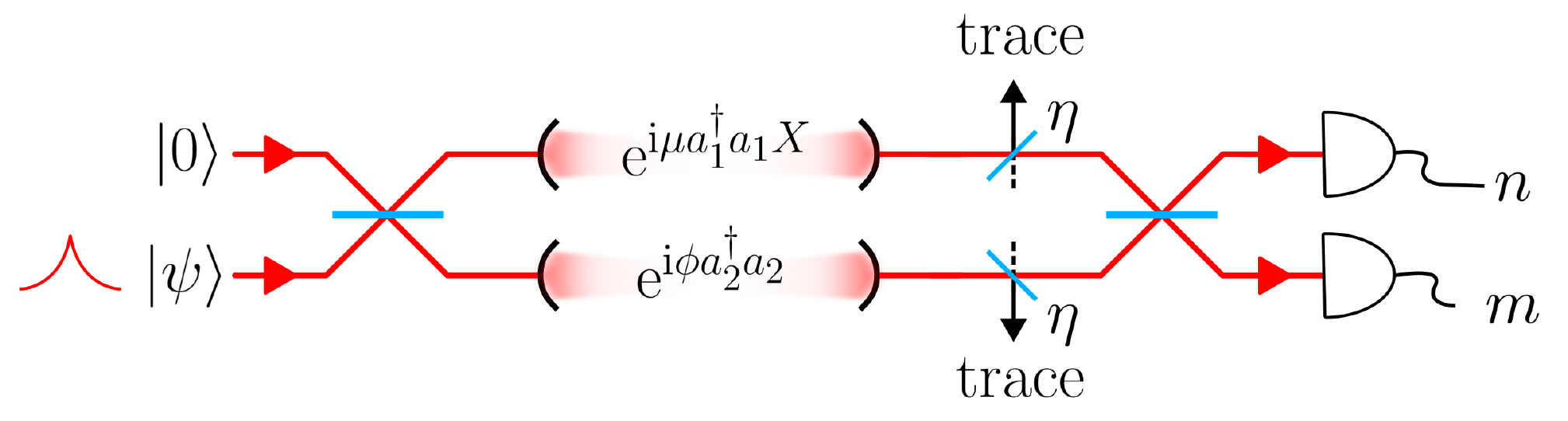} 
\caption{\small{{\textbf{Beam-splitter model for optical loss at any given step in the multistep protocol}}. Beam splitters of intensity transmission $\eta$ are placed in the interferometer path after the cavities to model loss to the optical environment. The input environmental modes interact with the interferometer modes on these beam splitters. A trace operation is then performed over the output environmental modes that leads to loss of optical information, which induces mechanical decoherence. At $\eta=1$, these beam splitters are fully transmissive and the protocol operates at perfect optical efficiency.}}
\label{losspdf}
\end{framed}
\end{figure}

In our multistep protocol, optical losses and detector inefficiencies are modelled by beam splitters of intensity transmission $\eta$ placed after the cavities in the upper and lower arms of the interferometer path, see Fig.~\ref{losspdf}. We introduce the field operators $a_{3}$ and $a_{4}$, which support the environmental modes impinging on these beam splitters in the upper and lower arms of the interferometer, respectively.
At optical frequencies the input environmental states are well described by the vacuum state and therefore the measurement operator corresponding to losing $k$ photons in the upper arm and $l$ photons in the lower arm, at any step in the protocol, is
\begin{eqnarray}
\Upsilon_{m,n,k,l}=\bra{m}\bra{n}\bra{k}\bra{l}U_{12}U_{24}U_{13}\mathrm{e}^{\rmi{\mu}a_{1}^{\dagger}a_{1}X}\mathrm{e}^{\rmi{\phi}a_{2}^{\dagger}a_{2}}U_{12}\ket{\psi}\ket{0}\ket{0}\ket{0}.
\end{eqnarray}
Here, the step index $j$ from the main text has been dropped for convenience. Subscripts have been included on the beam splitter unitaries to indicate the modes on which they act. $U_{12}$ operates on the interferometer modes as outlined in Section \ref{basicprotocol}, while $U_{13}$ and $U_{24}$ act according to
\begin{eqnarray}
U_{13}^{\dagger}a_{1}U_{13}=\sqrt{\eta}a_{1}+(\sqrt{1-\eta}~)a_{3},\\
U_{24}^{\dagger}a_{2}U_{24}=\sqrt{\eta}a_{2}+(\sqrt{1-\eta}~)a_{4}.
\end{eqnarray}
To account for the loss of information to the environment a trace operation must be performed over the output environmental states, so that after an $\{m,n\}$ click event the state of the mechanical mode is given by
\begin{eqnarray}\label{mapb}
\rho'=\frac{\sum_{k,l}\Upsilon_{m,n,k,l}\circ\rho}{\mathrm{tr}(\sum_{k,l}\Upsilon_{m,n,k,l}\circ\rho)}.
\end{eqnarray}
Here, $\rho$ is the state of the mechanical mode before the optomechanical interaction.

\subsection{Single photon input}
For a single photon input the measurement operator becomes
\begin{eqnarray}
\fl\Upsilon_{m,n,k,l}=\frac{1}{2}\sqrt{\eta}({\rm{e}}^{\rmi\mu{X}}+{\rm{e}}^{\rmi\phi}\mathds{1})\delta_{m,1}\delta_{n,0}\delta_{k,0}\delta_{l,0}+\frac{1}{2}\sqrt{\eta}({\rm{e}}^{\rmi\mu{X}}-{\rm{e}}^{\rmi\phi}\mathds{1})\delta_{m,0}\delta_{n,1}\delta_{k,0}\delta_{l,0}\nonumber\\
+\sqrt{\frac{1-\eta}{2}}{\rm{e}}^{\rmi\mu{X}}\delta_{m,0}\delta_{n,0}\delta_{k,1}\delta_{l,0}+\sqrt{\frac{1-\eta}{2}}{\rm{e}}^{\rmi\phi}\mathds{1}\delta_{m,0}\delta_{n,0}\delta_{k,0}\delta_{l,1}.
\end{eqnarray}
By inspection of $\Upsilon_{m,n,k,l}$ it is clear that for a $\{0,1\}$ or $\{1,0\}$ click event the only effect of optical loss is to reduce the heralding probability at each step by a factor of $\eta$. Importantly, the heralded mechanical state is not affected and in this way the protocol is resilient to optical loss and detector inefficiencies.

\subsection{Coherent state input}
When the input optical state is a coherent state $\ket{\sqrt{2}\alpha}$ the measurement operator takes the form
\begin{eqnarray}
\Upsilon_{m,n,k,l}=\frac{{\rm{e}}^{-|\alpha|^2}}{\sqrt{m!n!k!l!}}\frac{(\sqrt{\eta}\alpha)^{m+n}[(\sqrt{1-\eta})\alpha]^{k+l}}{(\sqrt{2})^{m+n}}\nonumber\\\times({\rm{e}}^{\rmi\mu X}+{\rm{e}}^{\rmi\phi}\mathds{1})^{m}({\rm{e}}^{\rmi\mu X}-{\rm{e}}^{\rmi\phi}\mathds{1})^{n}\mathrm{e}^{\rmi{k}\mu{X}}{\rm{e}}^{\rmi{l}\phi}\mathds{1}.
\end{eqnarray}
By considering the role of this operator in the map described by Eq.~(\ref{mapb}), we observe that the terms depending on the index $l$ have no operational effect on the mechanical mode and may be summed separately to give
\begin{eqnarray}
\sum_{l=0}^{\infty}\frac{|(\sqrt{1-\eta})\alpha|^{2l}}{l!}=\mathrm{e}^{(1-\eta)|\alpha|^2}.
\end{eqnarray}
The $k$-dependent terms affect the mechanical mode via
\begin{eqnarray}
\sum_{k=0}^{\infty}\frac{|(\sqrt{1-\eta})\alpha|^{2k}}{k!}\mathrm{e}^{\rmi{k}\mu{X}}\rho~\mathrm{e}^{-\rmi{k}\mu{X}},
\end{eqnarray}
which in the limit $(1-\eta)|\alpha|^2\ll1$ leaves the mechanical mode unchanged. In this regime, carrying out the trace operation leads to an effective measurement operator, for a $\{m,n\}$ photodetection event, given by
\begin{eqnarray}
\Upsilon_{m,n}=\frac{{\rm{e}}^{-\eta|\alpha|^2}}{\sqrt{m!n!}}\frac{(\sqrt{\eta}\alpha)^{m+n}}{(\sqrt{2})^{m+n}}({\rm{e}}^{\rmi\mu{X}}+{\rm{e}}^{\rmi\phi}\mathds{1})^{m}({\rm{e}}^{\rmi\mu{X}}-{\rm{e}}^{\rmi\phi}\mathds{1})^{n},
\end{eqnarray}
which is identical to Eq.~(\ref{kraussoperator}) aside from the transformation $\alpha\rightarrow\sqrt{\eta}\alpha$. In this regime the heralding probability is given by Eq.~(\ref{heralding}). 

When $(1-\eta)|\alpha|^2\ll1$ cannot be satisfied then the measurement operator takes the form 
\begin{eqnarray}
\Upsilon_{m,n,k}=\Upsilon_{m,n}\Theta_{k},
\end{eqnarray}
where $\Theta_{k}=C_{k}~D(\rmi{k}\mu/\sqrt{2})$, with $C_{k}=[(\sqrt{1-\eta})\alpha]^{k}/{\sqrt{k!}}$. Note that because $\Theta_{k}$ is a function of $X$, it commutes with all other operators in $\Upsilon_{m,n}$. As such, in the multistep protocol the effect of optical loss may be computed by commuting all of the $\Theta_{k}$ operators to left, which modifies Eq.~(\ref{rhoout2})  to
\begin{eqnarray}
\rho_{N}\propto\sum_{k_{1},k_{2},\ldots,k_{N}=0}^{\infty}\Theta_{k_{N}}\Theta_{k_{N-1}}\ldots\Theta_{k_{1}}[D(\rmi{N}\mu/\sqrt{2})-\mathds{1}]\circ\rho_{\mathrm{in}}.
\end{eqnarray}
In the case of a coherent input state and in the presence of optical loss, the Wigner distribution of the mechanical mode after $N$ steps of the protocol is therefore 
\begin{eqnarray}\label{wigloss}
 W_{\eta}(X,P)=\frac{\mathcal{N_{\eta}}}{\pi(1+2\bar{n})}\sum_{k_{1},k_{2},\ldots,k_{N}=0}^{\infty}|C_{k_{N}}|^2|C_{k_{N-1}}|^2\ldots|{C}_{k_{1}}|^2\nonumber\\
\Bigg\{{\exp}\Big[\frac{-X^2-(P-\mu\sum_{i=1}^{N}k_{i})^2}{1+2\bar{n}}\Big]+{\exp}\Big[\frac{-X^2-(P-N{\mu}-\mu\sum_{i=1}^{N}k_{i})^{2}}{1+2\bar{n}}\Big]\nonumber\\
-2{\rm{cos}}(N{\mu}X){\exp}\Big[\frac{-X^2-(P-N{\mu}/2-\mu\sum_{i=1}^{N}k_{i})^2}{1+2\bar{n}}\Big]\Bigg\},
\end{eqnarray}
where the normalization is given by
\begin{eqnarray}
\fl 1/2\mathcal{N}_{\eta}=\sum_{k_{1},k_{2},\ldots,k_{N}=0}^{\infty}|C_{k_{N}}|^2|C_{k_{N-1}}|^2\ldots|{C}_{k_{1}}|^2\Bigg\{1-\exp[-N^2\mu^2(1+2\bar{n})/4]\Bigg\}.
\end{eqnarray}
$W_{\eta}(X,P)$ describes a mechanical state consisting of a statistical mixture of SCSs described by the Wigner distribution $W_{\mathrm{SCS}}(X,P)$ from Eq.~(\ref{Wigscseq}). Hence, we may write $W_{\eta}(X,P)$ as 
\begin{eqnarray}\label{wigloss2}
\fl W_{\eta}(X,P)=\frac{\mathcal{N_{\eta}}}{\mathcal{N_{\mathrm{SCS}}}}\sum_{k_{1},k_{2},\ldots,k_{N}=0}^{\infty}|C_{k_{N}}|^2|C_{k_{N-1}}|^2\ldots|{C}_{k_{1}}|^2~W_{\mathrm{SCS}}\Big(X,P-\mu\sum_{i=1}^{N}k_{i}\Big).
\end{eqnarray}
The first term in this sum represents the mechanical state in the absence of loss, while higher order terms in $(1-\eta)|\alpha|^2$ represent states displaced from this by integer multiples of $\mu$ along the momentum axis. This incoherent mixing of states in the mechanical phase-space distribution is the manifestation of mechanical decoherence induced by optical loss. For an arbitrary $\alpha$ and $\eta$, the full heralding probability for state preparation in the presence of optical loss is given by 
\begin{eqnarray}
P_{N}=\sum_{k_{1},k_{2},\ldots,k_{N}=0}^{\infty}|C_{k_{N}}|^2|C_{k_{N-1}}|^2\ldots|{C}_{k_{1}}|^22^{1-N}\mathrm{e}^{-2N\eta|\alpha|^2}\eta^{N}|\alpha|^{2N}\nonumber\\
\quad\quad\quad\quad\quad\quad\Bigg\{1-\exp[-N^2\mu^2(1+2\bar{n})/4]\Bigg\},
\end{eqnarray}
which reduces the formula given in Eq.~(\ref{heralding}) in the limit $(1-\eta)|\alpha|^2\ll1$.

\begin{figure}[h]
\begin{framed}\centering
\includegraphics[scale=0.215, angle=270]{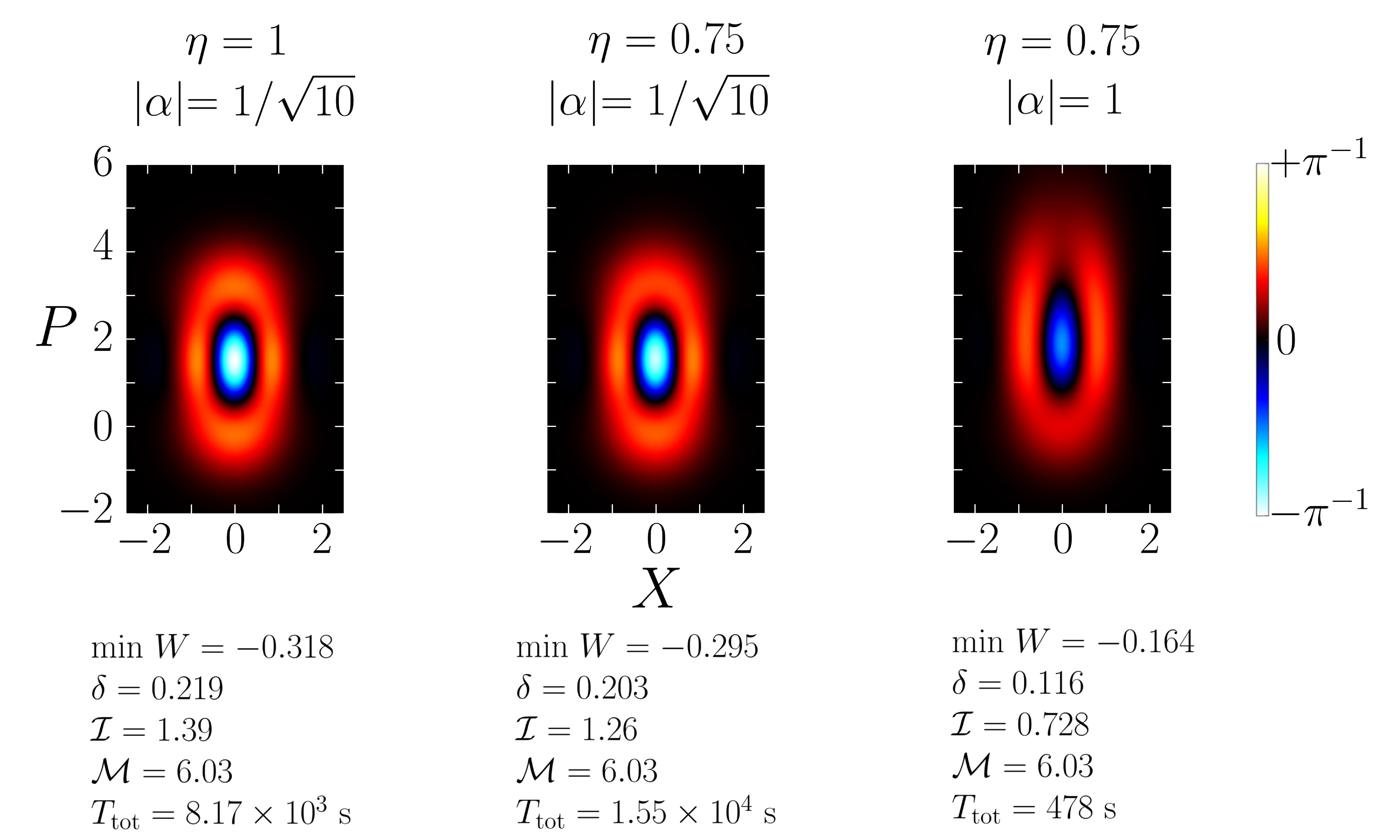} 
\caption{\small{{\textbf{The effect of optical loss and detector inefficiencies on the final mechanical state}}. In this figure we consider how optical loss affects the performance of a multistep protocol consisting of $N=3$ steps with a coherent state chosen as the optical input. We quantify this performance in terms of the non-classicality measures, the macroscopicity measures, and the total time taken to complete an experiment where one thousand statistical data points are obtained. We take the parameters from proposal (iii) of Table \ref{table1}, but ignore any thermal effects so that the effect of loss is isolated ($\mu=1$, $\bar{n}=0$). The phase-space plot at $\eta=1$ and $|\alpha|=1/\sqrt{10}$ (left) illustrates the case when the protocol operates at perfect optical efficiency. Conversely, the phase-space plot at $\eta=0.75$ and $|\alpha|=1/\sqrt{10}$ (middle) shows the effect that optical loss has in a regime where $(1-\eta)|\alpha|^2\ll1$. Here, the main effect of loss is to decrease $T_{\mathrm{tot}}$, while the phase-space measures are not changed significantly. The final plot illustrates a regime where optical loss leads to significant mechanical decoherence (right).}}
\label{lossboypdf}
\end{framed}
\end{figure}

The presence of optical loss also reduces the non-classicality and macroscopicity of the final mechanical state, however these changes can be made negligible for a given value of $\eta$ by reducing the amplitude of the input coherent state such that $(1-\eta)|\alpha|^2\ll1$ is satisfied. In Fig.~\ref{lossboypdf} we illustrate the effect of optical loss on the phase-space distribution of the heralded mechanical state. The asymmetry about $P=0$ in the final phase-space plot is explained by the interpretation of $W_{\eta}(X,P)$ as a statistical mixture of SCSs, each displaced by integer multiples of $\mu$ along the momentum axis. In this way we see that positive and negative areas of phase-space will combine in order to produce regions of reduced visibility as in the final phase-space plot of Fig.~\ref{lossboypdf}. Interestingly, Fig.~\ref{lossboypdf} shows that the macroscopicity measure $\mathcal{M}$ is unaffected by optical loss. This is because this measure depends entirely on the position probability marginal $p(X)$, and by integrating Eq.~(\ref{wigloss2}) we find that $p_{\eta}(X)=p_{\mathrm{SCS}}(X)$.

We would like to reiterate that the deleterious effects of optical loss to the mechanical state are avoided by choosing single photons or weak coherent states as the input states, as is considered in the main text.

\section{$\delta$ of a SCS} \label{appendixmax}
The negativity in $W_{\mathrm{SCS}}$ is due to the oscillating cosine term in Eq.~(\ref{Wigscseq}). For well-separated SCSs where $N\mu\gg1$, such that the overlap of the interference term with the population terms can be ignored, to calculate the value of $\delta_{\mathrm{SCS}}$ it is sufficient to consider only this term. From Eq.~(\ref{Wigscseq}) it is clear that any overlap of the interference term with the population terms will lead to a reduction in $\delta_{\mathrm{SCS}}$. The quantity 
\begin{equation}
\fl{E}=-2\frac{{\mathcal{N}}_{\mathrm{SCS}}}{(1+2\bar{n})\pi}\int_{-\infty}^{+\infty}\int_{-\infty}^{+\infty}\mathrm{d}X\mathrm{d}P{\rm{cos}}(N{\mu}X){\rm{exp}}\Big(\frac{-X^2-(P-N{\mu}/2)^2}{1+2\bar{n}}\Big)
\end{equation}
gives total volume introduced by the interference term. Note that $E<0$ because the minimum of the interference term occurs at $X=0$, $P=N\mu/2$ and away from this point the Wigner function is modulated by a decaying Gaussian envelope. $E$ may be decomposed as $E=\delta^{+}-\delta_{\mathrm{SCS}}$. Here, $\delta^{+}$ is the total positive volume and $\delta_{\mathrm{SCS}}$ is the non-classicality measure of interest in the limit $N\mu\gg1$. We may also consider the quantity $J$
\begin{equation}\label{Ieqn}
\fl{J}=2\frac{{\mathcal{N}}_{\mathrm{SCS}}}{(1+2\bar{n})\pi}\int_{-\infty}^{+\infty}\int_{-\infty}^{+\infty}\mathrm{d}X\mathrm{d}P|{\rm{cos}}(N{\mu}X)|{\rm{exp}}\Big(\frac{-X^2-(P-N{\mu}/2)^2}{1+2\bar{n}}\Big),
\end{equation}
which we may rewrite as $I=\delta^{+}+\delta_{\mathrm{SCS}}$. This therefore leads to the expression $\delta_{\mathrm{SCS}}=\frac{1}{2}(J-E)$. $E$ is readily calculated using basic Gaussian integration, and hence we obtain ${E}=-\exp[-N^2\mu^2(1+2\bar{n})/4]/\{1-\exp[-N^2\mu^2(1+2\bar{n})/4]\}$.
However, the absolute sign in Eq.~(\ref{Ieqn}) makes the evaluation of $J$ more involved. \par

Introducing the change of variable $\lambda=1/[N^2\mu^2(1+2\bar{n})]$ and $x=N{\mu}X$ allows us to write 
\begin{equation}
J=\frac{{2\mathcal{N}}_{\mathrm{SCS}}}{N\mu\sqrt{(1+2\bar{n})\pi}}I,
\end{equation}
with
\begin{eqnarray}
I=\int_{-\infty}^{+\infty}{\rm{e}}^{-\lambda{x}^{2}}|{\rm{cos}}(x)|\mathrm{d}x \nonumber \\
=\sum_{n=-\infty}^{\infty}\int_{-n\pi}^{(n+1)\pi}{\rm{e}}^{-\lambda{x}^{2}}|{\rm{cos}}(x)|\mathrm{d}x. 
\end{eqnarray}
By introducing the change of variables $t=x-n\pi$, $I$ becomes
\begin{equation}
\begin{array}{l}
I=\sum_{n=-\infty}^{\infty}\int_{0}^{\pi}{\rm{e}}^{-\lambda{(t+n\pi)}^{2}}|{\rm{cos}}(t)|\mathrm{d}t \nonumber \\
=\sum_{n=-\infty}^{\infty}\Bigg[\int_{0}^{\pi/2}{\rm{e}}^{-\lambda{(t+n\pi)}^{2}}{\rm{cos}}(t)\mathrm{d}t-\int_{\pi/2}^{\pi}{\rm{e}}^{-\lambda{(t+n\pi)}^{2}}{\rm{cos}}(t)\mathrm{d}t\Bigg]. \label{I7}
\end{array}
\end{equation}
Then in the second integral of Eq.~(\ref{I7}) we let $t\rightarrow{t}-\pi$, which allows $I$ to be written as
\begin{equation}
I= \sum_{n=-\infty}^{\infty}\int_{0}^{\pi/2}\Bigg({\rm{e}}^{-\lambda{(t+n\pi)}^{2}}+{\rm{e}}^{{-\lambda{((n+1)\pi-t)}}^{2}}\Bigg){\rm{cos}}(t)\mathrm{d}t. \label{I8}
\end{equation} 
The Poisson summation formula which relates the infinite sum over inverse Fourier components,
\begin{equation}
\sum_{n=-\infty}^{n=\infty}a(n)=\sum_{k=-\infty}^{k=\infty}\tilde{a}(k),
\end{equation}
may be used by assigning 
\begin{equation}
a(n)= \int_{0}^{\pi/2}\Bigg({\rm{e}}^{-\lambda{(t+n\pi)}^{2}}+{\rm{e}}^{{-\lambda{((n+1)\pi-t)}}^{2}}\Bigg){\rm{cos}}(t)\mathrm{d}t. 
\end{equation}
$\tilde{a}(k)$ is then calculated by Fourier transform,
\begin{equation}
\begin{array}{l}
\tilde{a}(k)=\int_{-\infty}^{+\infty}{\rm{e}}^{-2\pi{\rmi}ks}a(s)\mathrm{d}s\\
=\int_{0}^{\pi/2}\int_{-\infty}^{+\infty}{\rm{e}}^{-2\pi{\rmi}ks}\Bigg({\rm{e}}^{-\lambda{(t+n\pi)}^{2}}+{\rm{e}}^{{-\lambda{((n+1)\pi-t)}}^{2}}\Bigg){\rm{cos}}(t)\mathrm{d}t\mathrm{d}s \label{I11} \\ 
=\frac{2}{\sqrt{\lambda\pi}}{\rm{e}}^{-k^2/{\lambda}}\frac{(-1)^k}{1+(-1)^k2k}. 
\end{array}
\end{equation}
By putting Eq.~(\ref{I8})--(\ref{I11}) together, the value of $I$ is found to be 
\begin{equation}\label{lambda}
I=\sum_{k=-\infty}^{k=\infty}\frac{2}{\sqrt{\lambda\pi}}{\rm{e}}^{-k^2/{\lambda}}\frac{(-1)^k}{1+(-1)^k2k}.
\end{equation}

The total negative volume of the Wigner function $\delta_{\mathrm{SCS}}$ in the limit $N\mu\gg1$ is then given by
\begin{equation}
\fl\delta_{\mathrm{SCS}}=\frac{1}{2}\Big\{\frac{{4\mathcal{N}}_{\mathrm{SCS}}}{\pi}\sum_{k=-\infty}^{k=\infty}{\rm{e}}^{-k^2N^2\mu^2(1+2\bar{n})}\frac{(-1)^k}{1+(-1)^k2k} 
+\frac{\mathrm{e}^{-N^2\mu^2(1+2\bar{n})/4}}{[1-\mathrm{e}^{-N^2\mu^2(1+2\bar{n})/4}]}\Big\}. \label{deltalongeq}
\end{equation}
As $N\mu\rightarrow\infty$, Eq.~(\ref{deltalongeq}) tends to $1/\pi$ even in the case that the mechanical mode has a non-zero initial thermal occupation. Hence, by dropping the requirement that $N\mu\gg1$, we arrive at the result $0\leq\delta_{\mathrm{SCS}}<1/\pi$ valid for all finite $N$, $\mu$ and $\bar{n}$. Numerical results show that the introduction of thermal decoherence at each step in the protocol leads to a depletion in $\delta$ that cannot be compensated for by an increase in cat state separation. 

\section{The macroscopicity measure $\mathcal{M}$} \label{macroappendix}
The macroscopicity measure in bosonic systems for the quadrature degrees of freedom as introduced by Oudot \emph{et al} in Ref.~\cite{Oudot2015} is $N_{\mathrm{eff}}(\rho)=\frac{1}{2}\max\limits_{\theta,\mathrm{POVMs}}{F}_{\tau}$, where 
\begin{equation}
F_{\tau}=\int\mathrm{d}{x}\frac{1}{p_{\tau}(x)}\Big(\frac{\partial{p_{\tau}(x)}}{\partial{\tau}}\Big)^{2}. \label{CFIeqa}
\end{equation}
Here, $p_{\tau}(x)=\mathrm{tr}[\rho_{\tau}E(x)]$ for POVM element $E(x)$, $\rho_{\tau}=\mathrm{e}^{-\rmi\tau{X}_{\theta}}\rho\mathrm{e}^{\rmi\tau{X}_{\theta}}$, and $X_{\theta}=(a\mathrm{e}^{-\rmi\theta}+a^{\dagger}\mathrm{e}^{\rmi\theta})/\sqrt{2}$. As discussed in the main text, the difficult part in the calculation of $N_{\mathrm{eff}}(\rho)$ for a general state $\rho$ is the maximization over all possible POVMs. Hence we only consider a maximization over the complete set of quadrature measurements $E(X_{\lambda})=\ket{X_{\lambda}}\bra{X_{\lambda}}$, where $0\leq\lambda<\pi$. The macroscopicity measure we use in this work is therefore, $\mathcal{M}=\frac{1}{2}\max\limits_{\theta,\lambda}{F}_{\tau}.$

This expression can be simplified by rewriting the state $\rho_{\tau}$ in the Glauber-Sudarshan P respresentation,
\begin{eqnarray}
\rho_{\tau}=\mathrm{e}^{-\rmi\tau{X}_{\theta}}\rho\mathrm{e}^{\rmi\tau{X}_{\theta}} \nonumber \\
=\int_{-\infty}^{+\infty}\int_{-\infty}^{+\infty}P_{0}(X,P)\mathrm{e}^{-\rmi\tau{X}_{\theta}}\ket{\gamma}\bra{\gamma}\mathrm{e}^{\rmi\tau{X}_{\theta}}~\mathrm{d}^{2}\gamma \nonumber \\
=\int_{-\infty}^{+\infty}\int_{-\infty}^{+\infty}P_{\tau}(X,P)\ket{\gamma}\bra{\gamma}\mathrm{d}^{2}\gamma,
\end{eqnarray}
with coherent amplitude $\gamma=(X+iP)/\sqrt{2}$. Here $P_{\tau}(X,P)$ is the $P$ function of the state $\rho_{\tau}$, while $P_{0}(X,P)$ is the $P$ function of state $\rho$. Considering the action of the displacement operators on the coherent states $\ket{\gamma}$ allows one to see that $P_{\tau}(X,P)=P_{0}(X-\tau\sin\theta,P+\tau\cos\theta)$, and consequently $W_{\tau}(X,P)=W_{0}(X-\tau\sin\theta,P+\tau\cos\theta)$. The Radon transformation
\begin{equation}
p_{\tau}(X_{\lambda})=\int_{-\infty}^{+\infty}W_{\tau}(X\cos\lambda-P\sin\lambda,X\sin\lambda+P\cos\lambda)\mathrm{d}P
\end{equation}
may then be employed to show that $p_{\tau}(X_{\lambda})=p_{0}[X_{\lambda}-\tau\sin(\theta+\lambda)]$. Inserting this into Eq.~(\ref{CFIeqa}) and making the substitution $X_{\lambda}^{'}=X_{\lambda}-\tau\sin(\theta+\lambda)$ leads to the expression $F_{\tau}=\sin^2(\theta+\lambda)F_{X_{\lambda}}$. Here,
\begin{equation}
F_{X_{\lambda}}=\int_{-\infty}^{+\infty}\mathrm{d}{X_{\lambda}}\frac{1}{p_{0}(X_{\lambda})}\Big(\frac{\partial{p_{0}(X_{\lambda})}}{\partial{X_{\lambda}}}\Big)^{2},
\end{equation}
is the classical Fisher information of the quadrature $X_{\lambda}$, which is much easier to calculate than $F_{\tau}$ as no explicit reference is made to the phase-space translations which we use to test the sensitivity of the state $\rho$. Hence the macroscopicity measure we employ is given by $\mathcal{M}=\frac{1}{2}\max\limits_{\theta,\lambda}\{\sin^2(\theta+\lambda)F_{X_{\lambda}}\}$, which is clearly maximum when the phase-space translations are parallel to the quadrature $X_{\lambda}$, $\theta+\lambda=\pm\pi/2$. Finally we arrive at
\begin{equation}
\mathcal{M}=\frac{1}{2}\max\limits_{\lambda}{F}_{X_{\lambda}}.
\end{equation}

\section*{References}



\begin{thebibliography}{99}
%
\bibitem{GRW1986} Ghirardi G C, Rimini A, and Weber T 1986 \emph{Phys. Rev. D} \textbf{34} 470
%
\bibitem{Penrose1996} Penrose R 1996 \emph{Gen. Relat. Grav.} \textbf{28} 581
%
\bibitem{Diosi1989} Diosi L 1989 \emph{Phys. Rev. A} \textbf{40} 1165
%
\bibitem{BJK1999} Bose S, Jacobs K, and Knight P L 1999 \emph{Phys. Rev. A} \textbf{59} 3204
%
\bibitem{Marshall2003} Marshall W, Simon C, Penrose R, and Bouwmeester D 2003 \emph{Phys. Rev. Lett.} \textbf{91} 130401
%
\bibitem{Pikovski2012} Pikovski I, Vanner M R, Aspelmeyer M, Kim M S, and Brukner C 2012 \emph{Nat. Phys.} \textbf{8}, 393
%
\bibitem{Bosso2017} Bosso P, Das S, Pikovski I, and Vanner M R 2017\emph{Phys. Rev. A} \textbf{96}, 023849
%
\bibitem{Bose2017} Bose S, \emph{et al} 2017 \emph{Phys. Rev. Lett.} \textbf{119} 240401
%
\bibitem{Marletto2017} Marletto C, and Vedral V 2017 \emph{Phys. Rev. Lett.} \textbf{119} 240402
%
\bibitem{Bekenstein2012} Bekenstein J D 2012 \emph{Phys. Rev. D} \textbf{86} 124040
%
\bibitem{Haslinger2013} Haslinger P, Dorre N, Geyer P, Rodewald J, Nimmrichter S, and Arndt M 2013 \emph{Nat. Phys.} \textbf{9} 144
%
\bibitem{Clarke2008} Clarke J, and Wilhelm F K 2008 \emph{Nature} \textbf{453} 1031
%
\bibitem{Devoret2013} Devoret M, and Schoelkopf R J 2013 \emph{Science} \textbf{339} 1169
%
\bibitem{Berrada2013} Berrada T, van Frank S, Bucker R,Schumm T, Schaff J F and Schmiedmayer J 2013 \emph{Nat. Commun.} \textbf{4} 2077
%
\bibitem{RMP2014} Aspelmeyer M, Kippenberg T J, and Marquardt F 2014 \emph{Rev. Mod. Phys.} \textbf{86} 1391
%
\bibitem{Armour2002} Armour A D, Blencowe M P, and Schwab K C 2002 \emph{Phys. Rev. Lett.} \textbf{88} 148301
%
\bibitem{Akram2010} Akram U, Kiesel N, Aspelmeyer M, and Milburn G J 2010 \emph{New J. Phys.} \textbf{12} 083030
%
\bibitem{Khalili2010} Khalili F, Danilishin S, Miao H, Muller-Ebhardt H, Yang H, and Chen Y 2010 \emph{Phys. Rev. Lett.} \textbf{105} 070403
%
\bibitem{Marek2010} Marek P, and Filip R 2010 \emph{Phys. Rev. A} \textbf{81} 042325
%
\bibitem{Bennett2016} Bennett J S, Khosla K, Madsen L S, Vanner M R, Rubinsztein-Dunlop H, and Bowen W P 2016 \emph{New J. Phys.} \textbf{18} 053030
%
\bibitem{Vanner2011} Vanner M R \emph{et al} 2011 \emph{Proc. Natl. Acad. Sci.} \textbf{108} 16182
%
\bibitem{Lei2016} Lei C U \emph{et al} 2016 \emph{Phys. Rev. Lett.} \textbf{117} 100801
%
\bibitem{Romero2011} Romero-Isart O \emph{et al} 2011 \emph{Phys. Rev. Lett.} \textbf{107} 020405
%
\bibitem{VannerPRX} Vanner M R 2011 \emph{Phys. Rev. X} \textbf{1} 021011
%
\bibitem{Brawley2016} Brawley G A \emph{et al} 2016 \emph{Nat. Commun.} \textbf{7} 10988
%
\bibitem{Lee2012} Lee K C \emph{et al} 2012 \emph{Nat. Photon.} \textbf{6} 41
%
\bibitem{Vanner2013} Vanner M R, Aspelmeyer M, Kim M S 2013 \emph{Phys. Rev. Lett.} \textbf{110} 010504
%
\bibitem{Riedinger2016} Riedinger R \emph{et al} 2016 \emph{Nature} \textbf{530} 313
%
\bibitem{Milburn2016} Milburn T J, Kim M S, Vanner M R 2016 \emph{Phys. Rev. A} \textbf{93} 053818
%
\bibitem{Hoff2016} Hoff U B, Kollath-Bonig J, Neergaard-Nielsen J S, and Andersen U L 2016 \emph{Phys. Rev. Lett.} \textbf{117} 143601
%
\bibitem{Andrews2014} Andrews R W \emph{et al} 2014 \emph{Nat. Phys.} \textbf{10} 321
%
\bibitem{Tian2015} Tian L 2015 \emph{Annalen der Physik} \textbf{527} 1
%
\bibitem{Xia2014} Xia K, Vanner M R, and Twamley J 2014 \emph{Scientific Reports} \textbf{4} 5571
%
\bibitem{Rugar2004} Rugar D, Budakian R, Mamin H J, and Chui B W 2004 \emph{Nature} \textbf{430} 329
%
\bibitem{Hosseini2014} Hosseini M, Guccione G, Slatyer H J, Buchler B C, and Lam, P K 2014 \emph{Nat. Commun.} \textbf{5} 4663
%
\bibitem{Stannigel2012} Stannigel K \emph{et al} 2012 \emph{Phys. Rev. Lett.} \textbf{109} 013603
%
\bibitem{Ringbauer2016} Ringbauer M, Weinhold T J, Howard L A, White A G, and Vanner M R 2018 \emph{New J. of Phys.} \textbf{20} 053042
%
\bibitem{Kenfack2004} Kenfack A, and Zyczkowski K 2004 \emph{J. Opt. B: Quantum Semiclass. Opt.} \textbf{6} 396
%
\bibitem{Lee2011} Lee C-W, and Jeong H 2011 \emph{Phys. Rev. Lett.} \textbf{106} 220401
%
\bibitem{Frowis2018} Frowis F, Sekatski P, Dur W, Gisin N, and Sangouard N 2018 \emph{Rev. Mod. Phys.} \textbf{90} 025004
%
%
\bibitem{Brennecke2008} Brennecke F, Ritter S, Donner T, and Esslinger T 2008 \emph{Science} \textbf{322} 235
%
\bibitem{Purdy2010} Purdy T P \emph{et al} 2010 \emph{Phys. Rev. Lett.} \textbf{105} 133602
%
\bibitem{Neergaard-Nielsen2007} Neergaard-Nielsen J S, Nielsen B M, Takahashi H, Vistnes A I, and Polzik E S 2007 \emph{Optics Express} \textbf{15} 7940
%
\bibitem{Musslimani1995} Musslimani Z H, Braunstein S L,  Mann A, and Revzen M 1995  \emph{Phys. Rev. A} \textbf{51} 4967
%
\bibitem{Khosla2018} Khosla K E, Vanner M R, Ares N, and Laird E A 2018 \emph{Phys. Rev. X} \textbf{8} 021052
%
%
\bibitem{Braunstein1994} Braunstein S L and Caves C M 1994 \emph{Phys. Rev. Lett.} \textbf{72} 3439
%
\bibitem{Demkowicz-Dobrzanski2014}Demkowicz-Dobrzanski R, Jarzyna M, and Kolodynski K 2015 \emph{Prog. Opt.} \textbf{60} 345
%
\bibitem{Frowis2012} Frowis F and Dur W 2012 \emph{New J. Phys.}1\textbf{14} 093039
%
\bibitem{Oudot2015} Oudot E, Sekatski P, Frowis F, Gisin N,  and Sangouard N 2015 \emph{Journal of the Optical Society of America B} \textbf{32} 2190
%
\bibitem{Yadin2016} Yadin B, Vedral V 2016 \emph{Phys. Rev. A} \textbf{93} 022122 
%
\bibitem{Wilson2015} Wilson D J, Sudhir V, Piro N, Schilling R, Ghadimi A and Kippenberg T J 2015 \emph{Nature} \textbf{524} 325
%
\bibitem{Leijssen2017} Leijssen R, La Gala G R, Freisem L, Muhonen J T, and Verhagen E 2017 \emph{Nat. Commun.} \textbf{8} 16024
%
\bibitem{Arcizet2011} Arcizet O, Jacques V, Siria A, Poncharal P, Vincent P, and Seidelin S 2011 \emph{Nat. Phys.} \textbf{7} 879
%
\bibitem{Yin2013} Yin Z Q, Li T, Zhang X, and Duan L M 2013 \emph{Phys. Rev. A} \textbf{88} 033614
%
\bibitem{Lecocq2015} Lecocq F, Teufel J D, Aumentado J, and Simmonds R W 2015 \emph{Nat. Phys.} \textbf{11} 635
%
\end{thebibliography}
\end{document}